\documentclass[12pt]{article}
\usepackage{amssymb}
\usepackage{amsfonts}
\usepackage{amsmath,amsthm}
\usepackage{graphics,epsfig}
\usepackage{subfigure}
\usepackage{graphicx,epsfig, color }
\oddsidemargin 10mm \numberwithin{equation}{section}
\begin{document}
\begin{center}
{{\bf { Stabilization of spherically symmetric static Boson stars
by Coleman-Weinberg self-interaction potential}} \vskip 0.5 cm {
Firoozeh Eidizadeh \footnote{E-mail
address: firoozeh.eidizadeh@semnan.ac.ir} and Hossein Ghaffarnejad \footnote{E-mail address: hghafarnejad@semnan.ac.ir}}} \\
\vskip 0.1 cm {\textit{Faculty of Physics, Semnan University, P.C.
35131-19111, Semnan, Iran}}
\end{center}
\begin{abstract}
Our motivation in this work is due to the great importance of
relativistic boson stars as fierce competitors of black holes.
From theoretical perspective, they described by complex Klein
Gordon (KG) scalar fields interacting with curved background in
general form. In this work we use the Einstein-Klein Gordon (EKG)
scalar tensor gravity in presence of the Coleman-Weinberg (CW)
self-interaction potential to study formation of a spherically
symmetric boson star and situation of its stability. We choose the
CW potential because of its important role in description of
cosmic inflation.

\end{abstract}
\section{Introduction}
Ordinary stars are formed  by gas and dust clouds that are
distributed  non-uniformly throughout most galaxies in the
universe. All active ordinary stars eventually reach to a finite
scale object and collapses due to its own weight and it
 undergoes the process of stellar death \cite{sag}. This  process for most stars is responsible for the formation of very dense and compact
  stellar remnant, which is called compact (relativistic) star.
    In other words, compact stars, which includes
white dwarfs, neutron stars, quark stars, and black holes, are the
final stages in the evolution of ordinary stars. When an ordinary
star ceases its nuclear fuel supply, then its remnants takes one
of these relativistic forms, by depending on its mass during the
lifetime. One can follow \cite{Jit} for more discussions about
classification of the relativistic compact stars. Microscopically,
we can separate interior matter of relativistic compact stars to
two different kinds called as the bosonic and the fermionic
respectively. They have integer and half-integer spins and follow
different statistics laws, i.e., the Bose-Einstein and the
Fermi-Dirac statistical distribution respectively.
 However, the temperature is a dominant effect for the equation of
state (EOS) of ordinary stars, because of their molecular kinetic
energies, but that is not dominant effect for relativistic stars.
In the latter case, the supper-relativistic degenerate
bosons/fermions particles have energies in order of the fermi
energy for which the principle of exclusion of Pauli generate
degenerate pressure as dominant pressure independent of the
temperature. EOS for such a super-relativistic stars is a
relationship between the pressure and the density and the entropy.
One can see \cite{Jit} to find several kind of EOS used for
relativistic stars.  In summary, by 1939, investigation of
researchers dedicated on the problem of what happens to a compact
star core made entirely of degenerate fermions, i.e., the
electrons and the neutrons. Because of  these contributions, on
one side, nuclear matter and particle physics were becoming
essential for the description of matter at such extreme densities
and on the other hand, it had become clear that such superdense
objects could be described only within Einstein`s theory of
gravity. This caused to begin a new approach to world of
relativistic astrophysics. For studies in this direction, usually
the internal metric of a contracting star should be determined
with the help of Einstein's gravitational equation, which is known
as the Oppenheimer-Volkoff equation. This equation relates
pressure changes to the mass and density of matter inside the
star, which was first introduced by Robert Oppenheimer himself.
For instance, white dwarfs had provided in 1915, a new test of
Einstein's general theory of relativity \cite{con}.
\\
 A boson star is supported by a complex massive scalar field,
coupled to gravity given by general relativity or other
alternatives. There are two different approaches to investigate
formation of such stars, i.e., the classical and the quantum
fields theory approaches. In this work we use a classical approach
of the scalar field theory. In the classical regime of the
interacting matter fields to form a relativistic star, we can
choose one of two different approaches, namely time-dependent
evolution of collapsing bosonic cloud \cite{sco} or
time-independent one. We
 will work here within the context of classical
time-dependent field theory  for which the field has a complex
form (see for instance \cite{sco} \cite{MM4}, \cite{MMM4}). In
summary, the boson stars are particle-like solutions of the
Einstein metric equations that were found in the late 1960s. Since
then, boson stars have been used in a wide variety of models as
sources of dark matter, as black hole mimickers, in simple models
of binary systems, and as a tool for finding black holes in higher
dimensions. A full review for important varieties of boson stars,
their dynamic properties, and some of their applications,
concentrating on recent efforts, is collected in the ref.
\cite{stev}. Boson stars for a free scalar field without any kind
of self interaction potential was studied previously to obtain
boson star with maximum mass $M\approx M_{Pl}^2/m_b$, which is
less than the Chandrasekhar mass $M_{Ch}\approx M_{Pl}/m_f^2$
obtained from fermionic stars, and hence, they are called as
`mini` boson stars. In the latter formula, $m_b$ and $m_f$ are the
masses of the candidates particles of bosons and fermions,
respectively, and $M_{Pl}$ is the Planck mass. In order to extend
this limit and reach astrophysical masses comparable to the
Chandrasekhar mass, some potentials were generalized to include a
self-interaction term that provides an extra pressure against
 the gravitational collapse \cite{colp}. As another application of the EKG theory in presence of self interaction potentials as $\phi^6$ together with $\phi^4$ is used in ref
 \cite{M1} to investigate a  rotating boson star.
 Instead of using different forms of self-interacting potentials of scalar fields other physical effects, for instance, the electric field or
 magnetic field effects (see for instance \cite{Ghf} and \cite{Ley}), have also been studied for the investigation of the stability of boson stars.
 To follow this, one can see \cite{vic}, in which authors solved the Einstein-Maxwell-Klein-Gordon equations to obtain
  a compact, static axially symmetric magnetized object which
is electrically neutral.  It is composed from two complex massive
charged scalar fields and has several particular properties,
including the torus form of the matter density and the expected
dipolar distribution of the magnetic field, with some peculiar
features in the central regions. The authors showed that their
stellar model is free of divergencies in any of the field and
metric functions. Also they presented a discussion about their
model where the gravitational and magnetic fields in the external
region are similar to those of neutron stars. To study conditions
of stability of a boson star from the EKG equations interacting
with a perfect fluid matter, one can see \cite{Font}. Another
application of the EKG gravity model is presented in ref.
\cite{Hua} where its authors studied the properties of
Bose-Einstein Condensate (BEC) systems consisting of two scalar
fields, focusing on both scale of the stellar and the galactic
objects. They showed that the presence of extra scalars and
possible interactions between them can leave unique imprints on
the BEC system mass profile, especially when dominance of one
scalar is changed to the other. At stellar scales (nonlinear
regime), they presented that a repulsive interaction between the
two scalars of the type $+\phi_1^2\phi_2^2$ can stabilize the BEC
system and support it up to high compactness, a phenomenon  known
to exist only in the
 $\phi^4$ system. They provided  a simple analytic understanding of this behavior
and pointed out that it can lead to interesting gravitational wave
signals at LIGO-Virgo. At galactic scales, on the other hand, they
showed that two-scalar BECs can address the scaling problem that
arises when one uses ultralight dark matter mass profiles to fit
observed galactic core mass profiles. They constructed a particle
model of two ultralight scalars with the repulsive
$\phi_1^2\phi_2^2$ interaction using collective symmetry breaking.
\\
So far, two incredibly significant experimental results have
appeared in relation to the existence of boson stars: (a) The
first, scalar particle is so called the Higgs particle has been
found by the LHC,  although its instability causes that it dose
not considered as the fundamental constituent of boson stars. (b)
Far from the quantum particle regime of the LHC, the LIGO-Virgo
collaboration directly detected gravitational waves in 2015, which
were completely consistent with the merger of a binary black hole
system{\color{blue},} as predicted previously by general
relativity \cite{colp}. Hence, future work on boson stars may be
experimental more, particularly if fundamental scalar fields can
be observed or if evidence arises indicating that the boson stars
uniquely fit galactic dark matter. But regardless of any
experimental results found by these remarkable experiments, there
will always be unexplored regimes by experiments where boson stars
will find a natural place. Boson stars have a long history as
candidates for all manner of phenomena, from fundamental particle,
to galactic dark matter. A huge variety of solutions have been
found and their dynamics  have been studied. Mathematically, boson
stars  are fascinating soliton-like solutions. Astrophysically,
they represent possible explanations of black hole candidates and
dark matter, with observations constraining properties of boson
stars. Therefore, we would like to use in this paper, an EKG
scalar tensor gravity model in the presence of a CW potential
\cite{wei} and we investigate that how could formed a CW boson
star such that remain as stable. As we mentioned in the abstract
section our motivation comes from importance of such a potential
in rate of cosmic inflation. Usually Boson stars have not sharp
surface same as ordinary or neutron stars which their radius are
determined by setting with zero value the matter density and the
radial pressure. Density and pressures asymptotically vanish at
infinite distance usually. One of the merits of this paper is that
it shows that the CW potential ensures that the boson particles in
such a boson star do not decay quickly and thus the CW boson star
have a hard surface with finite radius.
 Layout of the paper is organized as follows.\\
 In section 2, we define the EKG gravity model and present the corresponding field equations.
 In section 3, we generate the explicit form
 of the equations for
  a spherically symmetric static curved line element. We find linear
  order solutions of the nonlinear equations of the fields.  This is done via dynamical system approach. We determined the
  critical points of the fields equations and calculated Jacobi matrix of the fields equations in these critical points.
   Then we solved secular equation of
  the Jacobi matrix and at last interpret situations where the system remain as  stable.
   In section 4, we interpret our obtained solutions and we plotted figures of physical quantities versus the radius and density parameters
    of the CW boson star.
   Mathematical derivations show that this is the boson mass which controls values of the star radius but in presence of the CW self-interaction potential.
 The summary and outlook of this work are dedicated to the last section.
  \section{The gravity model}
  As we pointed out above, in the case of boson stars, the
energy-momentum content is that of a complex valued scalar field
and so let`s start with EKG scalar tensor gravity with a complex
form for the scalar KG field such that
\begin{equation}
I=\frac{1}{16\pi}\int d^4 x \sqrt{g}[R-\frac{1}{2}g^{\mu
\nu}(\partial_\mu \psi^*
\partial_\nu \psi+\partial_\mu
\psi
\partial_\nu \psi^*)-m^2 \psi\psi^* -V(|\psi|)]
\end{equation}
  where $g$ is absolute value of determinant of the metric field $g_{\mu \nu}$
, $R$ is Ricci scalar and $m$ is mass parameter of the complex KG
scalar field $\psi$ (with complex conjugate $\psi^*$ and norm
$|\psi|=\sqrt{\psi\psi^*}$). $V(|\psi|)$ is arbitrary
self-interaction potential of the KG field. We use geometric
unites $c=1=G$ with metric signature $(+,-,-,-)$ in which $\psi$
is dimensionless, but $m$ has inverse of the length dimension and
the dimensions for self interaction potential $V(|\psi|)$ is
$(length)^{-2}$ too. To find the metric field of a boson star we
should first choose a particular form for $V(|\psi|).$  Several
types of boson stars, along with the definitions of their
corresponding potentials, are presented in the article \cite{M1}.
We choose here the CW potential and consider its effect on the
stability of the boson star.
At first,  we present a short review about the historical importance of the CW potential as follows:\\
Historically, the Higgs potential $V(|\psi|)=\lambda|\psi|^4$ with
coupling constant $\lambda>0$ was used to describe the chaotic
inflation in cosmology. Ordinarily, $\lambda<0$ is forbidden on
the grounds that it leads to a potential without a lower bound but
of course, this statement pertains only to the zero Feynman loop
approximation of the effective potential, i.e., the loop
contributions may (or may not) provide a lower bound on the
effective potential even for negative $\lambda$ \cite{wei}. It is
obvious that the above action functional contains a discrete
symmetry under the transformation of $\psi\to-\psi$ for the Higgs
potential $\lambda|\psi|^4$ \cite{Hig}. One may ask a question
such that how this symmetry can be broken in both the classical
and the quantum regimes of the KG field by modifying the potential
$\lambda|\psi|^4$? It is easy to check that the vacuum expectation
value of the KG field is found at the extremum of the potential
$m^2|\psi|^2+\lambda|\psi|^4$. This means that in the vacuum state
the fluctuations of the KG field should vanish. The vacuum
expectation value occurs at $<|\psi|>_{vac}=0$ which is symmetric
for $m^2>0$ (the real physical particles) but is anti-symmetric
$<|\psi|>_{vac}=\pm\sqrt{-m^2/2\lambda}$ for $m^2<0$ (the
tachyonic non-physical particles). In other words, if the mass
term is tachyonic, $m^{2}<0$ there is a spontaneous breaking of
the gauge symmetry at low energies, a variant of the Higgs
mechanism. On the other hand, if the squared mass is $m^{2}\geq0$
the vacuum expectation of the field $\psi$ is zero and the
symmetry breaking
is not occur.\\
To give a real perspective about the symmetry breaking, Erick
Weinberg and his supervisor Sidney Coleman demonstrated \cite{wei}
that even if the re-normalized mass is zero, the spontaneous
symmetry breaking still happens due to the radiative corrections
from at least one-loop Feynman diagrams in the interacting KG
field. In short, they showed that radiative corrections to mass
re-normalized effective potential generates a logarithmic singular
term in the effective potential called as the generalized CW
potentials such that
\begin{equation}\label{cw}V_{cw}(|\psi|)=\lambda\bigg[|\psi|^4\ln\bigg(\frac{|\psi|}{|\psi_0|}\bigg)-\frac{1}{4}|\psi|^4+\frac{1}{4}|\psi_0|^4\bigg]\end{equation}
where in the geometric units $c=G=1$ the dimensions for the
coupling constant $\lambda$ is $(length)^{-2}$ because $|\psi|$ is
dimensionless. This model introduces a mass scale $|\psi_0|$ for a
classically conformal theory of the model with a conformal
anomaly. In fact $|\psi_0|$ is originated from ultraviolet cutoff
frequency in calculating the Feynman path integrals of radiative
corrections in one loop level which should be re-normalized. This
in fact describes that how  massive (Goldstone) gauge bosons can
be created from the self-interaction of elementary massless bosons
which is called as the Higgs mechanisms (related to the
Goldstone`s theorem \cite{dan}).
 The potential (\ref{cw}) is applicable in the new inflation in
 cosmic models because the potential is very flat and has a maximum value at
 $|\psi|=0$ (see figure 1-a). The scalar field dose not escape from classical tunneling
 via Sphalerons at high temperature, but due to quantum fluctuations via Instantons at low temperatures, (see chapter four in ref. \cite{Mukh}).
This kind of potential describes  self-reproduction of the
universe and primordial inhomogeneities of the universe
successfully but, after the end of the primordial inflation
\cite{Mukh}. To see importance of the CW potential more, one can
follow other references for instance \cite{M2}, \cite{M3},
\cite{M4}, \cite{M5},\cite{M7}. Despite the usefulness of this
type of potential in theoretical studies of cosmic inflation,
there are some inconsistencies with observational data from Planck
2018 and even DESI2024 datasets. For instance the tensor-to-scalar
ratio in CW inflation dose not satisfied the observational date
(see page 6 in ref. \cite{Liao}). However, we like to investigate at below, the physical effects of the CW potential (\ref{cw}) on
formation and stability of a boson star in presence of fluctuations of massive bosons.\\
By varying the above action functional with respect to the fields
$\psi^*$ and $g^{\mu\nu}$ we obtain equations of motion for the KG
field $\psi$ and the metric field respectively such that
\begin{align} \label{KG}
\Box \psi-
m^2\psi-2\lambda(\psi\psi^*)\psi\ln\bigg(\frac{\psi\psi^*}{\psi_0\psi^*_0}\bigg)=0,~~~~\Box=g^{-\frac{1}{2}}\partial_{\mu}(g^\frac{1}{2}g^{\mu\nu}\partial_\nu)
\end{align}
and \begin{align}\label{Ein} G_{\mu \nu}=T_{\mu \nu}(|\psi|),
\end{align}
where
\begin{align}\label{Tmunu}
T_{\mu \nu}(|\psi|)&=\frac{1}{2}(\partial_\mu \psi^*
\partial_\nu \psi+\partial_\mu \psi \partial_\nu
\psi^*)-\frac{g_{\mu \nu}}{2} \bigg\{\frac{1}{2}g^{\alpha
\beta}(\partial_\alpha \psi^*
\partial_\beta \psi+\partial_\alpha \psi \partial_\beta
\psi^*)\notag\\&+m^2\psi\psi^*+\lambda\bigg[\frac{(\psi\psi^*)^2}{2}\ln\bigg(\frac{\psi\psi^*}{\psi_0\psi^*_0}\bigg)
-\frac{1}{4}(\psi\psi^*)^2+\frac{1}{4}(\psi_0\psi^*_0)^2\bigg]\bigg\}.
\end{align}
 We should remember that this model contains
an additional conserved Noether current
\begin{align} j^\mu=-\frac{i}{2}(\psi^*\nabla^\mu\psi-\psi\nabla^\mu\psi^*),~~~\nabla_\mu j^\mu=0\end{align} due to the internal global $U(1)$ symmetry
$\psi\to e^{i\chi}\psi$ in which $\chi$ is a constant field. In
the subsequent section, we set these equations for a spherically
symmetric static curved line element to investigate formation of a
boson star.
\section{Coleman-Weinberg boson star }
Let's start with the following ansatz for the spherically
symmetric complex KG scalar field which makes a boson star
\cite{Betti}
\begin{equation}\label{psitr}\psi(t,r)=e^{i\omega
t}\phi(r),~~~~\omega>0.\end{equation} We assume that the above KG
scalar field participates in the self interaction CW potential
(\ref{cw}) for which the isotropic spherically symmetric curved
space-time line element is given in the standard form by (see for
instance \cite{Wein} chapter
8)\begin{align}\label{line1}ds^2=e^{U(r)}dt^2-
e^{H(r)}dr^2-r^2(d\theta^2+\sin^2\theta d\varphi^2).\end{align}
Substituting (\ref{psitr}), the KG wave equation (\ref{KG}) reads
\begin{equation}\label{KG11}\phi^{\prime\prime}+\bigg(\frac{2}{r}+\frac{U^\prime}{2}-\frac{H^\prime}{2}\bigg)\phi^\prime+e^H\bigg[(\omega^2
e^{-U}+m^2)\phi+4 \lambda
\phi^3\ln\bigg(\frac{\phi}{\mu}\bigg)\bigg]=0
\end{equation}
where $\prime$ is derivative with respect to $r$ coordinate, the
constant parameter $\mu=|\psi_0|$ comes from ultraviolet cut off
scales when we do evaluate re-normalized expectation values of the
quantum KG field $\psi$ in presence of the radiative corrections
of the Feynman diagrams. Substituting the line element
(\ref{line1}) into the Einstein equation (\ref{Ein}) we find
\begin{align}\label{tt1}&\frac{H^\prime}{r}-\frac{1-e^{H}}{r^2}+\frac{\phi^{\prime2}}{2}+\frac{\omega^2\phi^2e^{H-U}}{2}\notag\\&
-\frac{e^{H}}{2}\bigg\{m^2\phi^2+\lambda\bigg[\phi^4\ln\bigg(\frac{\phi}{\mu}\bigg)-\frac{1}{4}\phi^4+\frac{1}{4}\mu^4\bigg]\bigg\}=0
\end{align} for tt component
\begin{align}\label{rr1}\frac{U^\prime}{r}&+\frac{\phi^{\prime2}}{2}+\frac{1-e^{H}}{r^2}+\frac{\omega^2\phi^2e^{H-U}}{2}
\notag\\&+\frac{e^{H}}{2}\bigg\{m^2\phi^2+\lambda\bigg[\phi^4\ln\bigg(\frac{\phi}{\mu}
\bigg)-\frac{\phi^4}{4}+\frac{\mu^4}{4}\bigg]\bigg\}=0\end{align}
for rr component and
\begin{align}\label{thth1}&U^{\prime\prime}+\frac{U^{\prime2}}{2}-\frac{H^\prime U^\prime}{2}+\frac{U^\prime}{r}-\frac{H^\prime}{r}
-\phi^{\prime2}+\omega^2\phi^2e^{H-U}\notag\\&+e^{H}\bigg[m^2\phi^2+\lambda
\bigg[\phi^4\ln\bigg(\frac{\phi}{\mu}\bigg)-\frac{1}{4}\phi^4+\frac{1}{4}\mu^4\bigg]\bigg]=0\end{align}
for angular components. In the above equations we used the density
$\rho(r),$ the radial pressure $p_r(r)$ and the transverse
pressure $p_t(r)$ of the KG field together with the CW potential
(\ref{cw}) such that
\begin{align} \label{Tmat}&\rho=T^t_t=\frac{e^{-H}\phi^{\prime2}}{2}+\frac{e^{-U}\omega^2\phi^2}{2}-\frac{1}{2}\bigg\{m^2\phi^2
+\lambda\bigg[\phi^4\ln\bigg(\frac{\phi}{\mu}\bigg)-\frac{1}{4}\phi^4
+\frac{1}{4}\mu^4\bigg]\bigg\}\notag\\&
p_r=T^r_r=p_t-e^{-H}\phi^{\prime2}\notag\\&
p_t=T^\theta_\theta=T^\varphi_\varphi=\rho+\omega^2\phi^{2}e^{-U}.\end{align}
In fact, to have explicit from of the fields solutions $\phi(r)$,
$U(r)$ and $H(r)$ we need three equations from (\ref{KG11}),
(\ref{tt1}), (\ref{rr1}) and (\ref{thth1}) only and so one of them
is a constraint condition between the solutions. This claim is
proved via conservation equation of the stress tensor of the
gravitational system or equivalently the Bianchi identity in which
related three components of the Einstein metric equations
(\ref{tt1}), (\ref{rr1}) and (\ref{thth1}) to each other. The
covariant conservation of matter stress tensor or Bianchi`s
identity $\nabla_\mu G_\nu^\mu=\nabla_\mu T_\nu^\mu=0$ defined by
internal metric of the stellar object is usually called as
Tolman-Oppenheimer-Volkoff (TOV) equation (see for instance
\cite{cat} and \cite{Hob}) which for the line element
(\ref{line1}) reads to the following form
\begin{equation}\label{TOV1}p_r^\prime-\frac{U^\prime}{2}\rho+\bigg(\frac{U^\prime}{2}+\frac{2}{r}\bigg)p_r-\frac{2p_t}{r}=0\end{equation}
where  $\rho(r),p_r(r)$ and $p_t(r)$ should be substituted by
(\ref{Tmat}). If there is determined form of the equation of state
of the gravitational system then one can solve the equation of TOV
instead of the Einstein field equations. The equation of state is
a relationship between the directional pressures $p_{r,t}$ and the
density function $\rho$. Although in the literature some of
equation of states are presented for relativistic stars but we do
not use them same as our previous work \cite{Ghf} and we like to
be free in this work such that after to solve the metric equations
then we find explicit form by eliminating the radius parameter
between (\ref{Tmat}) (see figures 3, 4). Thus
 to determine explicit form of the
functions $U(r), H(r)$ and $\phi(r)$ we choose (\ref{KG11}) and
two new generated equations $(\ref{tt1})+(\ref{rr1})=0$ and
$2(\ref{tt1})+(\ref{thth1})=0$ which in a dimensionless forms they
are respectively
\begin{align}\label{311}\ddot{\sigma}+\dot{\sigma}^2+\bigg(\frac{2}{\tau}+\frac{\dot{U}}{2}-\frac{\dot{H}}{2}\bigg)\dot{\sigma}
+e^H\bigg[\bar{\omega}^2 e^{-U}+\bar{m}^2+4\epsilon\mu^2\sigma
e^{2\sigma}\bigg]=0,\end{align}
\begin{align}\label{312}\dot{H}=-\dot{U}-\tau\mu^2 e^{2\sigma}(\dot{\sigma}^2+\bar{\omega}^2 e^{H-U})\end{align} and
 \begin{align}\label{313}\ddot{U}+\frac{\dot{U}^{2}}{2}-\frac{\dot{H}\dot{U}}{2}+\frac{\dot{H}+\dot{U}}{\tau}-\frac{2(1-e^{H})}{\tau^2}+2\bar{\omega}^2\mu^2 e^{2\sigma+H-U}=0\end{align}
  in which we defined dimensionless quantities as
 \begin{align}\sigma=\ln(\phi/\mu),~~~\tau=\frac{r}{\ell},~~~&\epsilon=\lambda\ell^2,~~~\bar{m}=m\ell,~~~\notag\\&\bar{\omega}=\omega\ell,~~~\prime
 =\frac{d}{dr}=\frac{1}{\ell}\frac{d}{d\tau}=\cdot.\end{align} The equations (\ref{311})
 , (\ref{312}) and (\ref{313}) are
nonlinear second order differential equations and to solve them we
are free to choose two different ways, i.e., the numeric or the
analytic perturbation series methods. We use perturbation series
method to find analytic solutions near the asymptotic surfaces
$\tau\to0$ and $\tau\to\infty$. To do this, we use dynamical
system approach where each of higher order differential equation
transforms to several differential equations with first order by
defining new fields such that
\begin{equation}\label{xy}\dot{\sigma}=x,~~~\dot{U}=y.\end{equation}
Substituting (\ref{xy}) and $\dot{H}$ given by (\ref{312})
 the equations (\ref{311}), (\ref{312}) and (\ref{313})
can be rewritten  respectively as follows
\begin{align}\label{xdot}&\dot{x}=-x^2-x[2/\tau+(\tau/2)\mu^2
e^{2\sigma}(x^2+\bar{\omega^2} e^{H-U})]\notag\\&~~~~~
-\bar{\omega}^2 e^{H-U}-\bar{m}^2 e^{H}-4\epsilon \mu^2 \sigma
e^{H+2\sigma}\notag\\& \dot{H}=-y-\tau
\mu^2e^{2\sigma}(x^2+\bar{\omega}^2 e^{H-U})\notag\\&
\dot{y}=-y^2-(\tau/2)\mu^2 y e^{2\sigma}(x^2-\bar{\omega}^2
e^{H-U})+\mu^2 e^{2\sigma}(x^2+\bar{\omega}^2
e^{H-U})\notag\\&~~~~~ +(2/\tau^2)(1-e^H)-2\bar{\omega}^2\mu^2
e^{H-U+2\sigma}.\end{align} From point of the dynamical system
approach, the set of variables $\{x,H,y,\sigma,U\}$ make 5
dimensional phase space which satisfies the 5 nonlinear
differential equations (\ref{xy}) and (\ref{xdot}).  To solve
these equations with a dynamical system approach one usually
should expand them around the critical points in
 phase space which are obtained from the equations \begin{equation}\label{crit}
 \dot{x}=0,~~~\dot{H}=0,~~~\dot{y}=0,~~~\dot{\sigma}=0,~~~\dot{U}=0
\end{equation} and then keep up linear order perturbation series expansion of the equations (\ref{xy}) and (\ref{xdot}) near the critical points
which their coefficients are called usually as the Jacobi matrix.
By determining sign of eigenvalues of this Jacobi matrix one can
find nature of stability of the obtained solutions and also by
solving the linear order equations one can obtain asymptotic
solutions of the fields near the  critical points (see suitable
text books for instance \cite{lav} or introduction section of the
ref. \cite{GHY} for more descriptions about the dynamical system
approach). After to present a little description about the
dynamical system approach we are now ready to calculate the
critical points of the equations (\ref{xy}) and (\ref{xdot}).  In
this case, the equations (\ref{crit}) give the following values
for the critical points
\begin{align}
P_c=\{x_c=0,~~~y_c=0,~~~
H_c=0,~~~e^{-U_c}=-\frac{\bar{m}^2}{\bar{\omega}^2},~~~\sigma_c=-\infty(\phi_c=\mu)\}.\end{align}
Elements of the Jacobi matrix on the critical point above are
obtained as
\begin{align}J_{ij}|_{P_c}=\frac{\partial \dot{x}_i}{\partial x_i}\big|_{P_c}=\left(%
\begin{array}{ccccc}
  -2/\tau & 0 & 0 & 0& -\bar{m}^2 \\
  0 & 0 & -1 & 0 & 0 \\
  0 & -2/\tau^2 & 0 & 0 & 0 \\
  1 & 0 & 0 & 0 & 0 \\
  0 & 0 & 1 & 0 & 0 \\
\end{array}%
\right)\end{align}\\
where we consider the asymptotic surface $\tau\to0$ because a
boson star is a compact object and so its matter is localized. As
an extension of the work one can seek same calculates for regimes
$\tau\to\infty$ where the boson particles make a cloud
distribution. However one can solve secular equation
$\det({J_{ij}-\delta_{ij} E})=0$ of the Jacobi matrix above to
find eigenvalues $E$ as
\begin{equation}E^2(E+\frac{2}{\tau})(E^2-\frac{2}{\tau^2})=0\end{equation} with solutions \begin{equation}E_{1,2}=0,~~~E_3=-\frac{2}{\tau},~~~E_{4,5}=\pm\frac{\sqrt{2}}{\tau}
\end{equation} where $\tau>0$ and so positive values of the eigenvalues above show instability nature of the
system but negative values show stable nature. $E_{1,2}$ with zero
values show that the system is degenerate in stabilization and to
make stable more we should consider other interaction potentials
such that these degeneracy break to make negative eigenvalues more
than. At present form the obtained solutions show quasi-stable
nature of the system because some of the eigenvalues have negative
sign but some of them have positive sign (see figure 1-b). Linear
order form of the equations (\ref{xy}) and (\ref{xdot}) read
\begin{equation}\frac{d}{d\tau}\left(%
\begin{array}{c}
  x \\
  H \\
  y\\
  \sigma \\
 U \\
\end{array}%
\right)=\left(%
\begin{array}{ccccc}
  -2/\tau & 0 & 0 & 0& -\bar{m}^2 \\
  0 & 0 & -1 & 0 & 0 \\
  0 & -2/\tau^2 & 0 & 0 & 0 \\
  1 & 0 & 0 & 0 & 0 \\
  0 & 0 & 1 & 0 & 0 \\
\end{array}%
\right)\left(%
\begin{array}{c}
  x \\
  H \\
  y\\
  \sigma \\
 U \\
\end{array}%
\right).\end{equation}  It is not complicated to find solutions of
the above matrix equation as \begin{equation}\left(%
\begin{array}{c}
  x \\
  H \\
  y\\
  \sigma-\sigma_c \\
 U-U_c \\
\end{array}%
\right)=\tau^\eta\left(%
\begin{array}{c}
  \bar{m}^2\tau/(3+\eta) \\
  1 \\
  -\eta/\tau\\
  \bar{m}^2\tau^2/(2+\eta)(3+\eta) \\
 -1 \\
\end{array}%
\right),~~~\eta_{\pm}=\frac{1\pm3}{2}\end{equation} for which we
obtain
\begin{align}\label{metsol}&\phi=\mu\exp\bigg[\frac{\bar{m}^2\tau^{2+\eta}}{(2+\eta)(3+\eta)}\bigg],~~~ e^U=-\frac{\bar{\omega}^2}{\bar{m}^2}\exp(-\tau^\eta),~~~e^{H}=\exp(\tau^\eta)\notag\\&
\phi^+=\mu\exp\bigg[\frac{\bar{m}^2\tau^{4}}{20}\bigg],~~~
e^{U_+}=-\frac{\bar{\omega}^2}{\bar{m}^2}\exp(-\tau^2),~~~e^{H_+}=\exp(\tau^2)\notag\\&
\phi^-=\mu\exp\bigg[\frac{\bar{m}^2\tau}{2}\bigg],~~~
e^{U_-}=-\frac{\bar{\omega}^2}{\bar{m}^2}\exp(-\tau^{-1}),~~~e^{H_-}=\exp(\tau^{-1})
\end{align}
which show two different kind of solutions. However, by
substituting these two different branches of solutions into the
stress tenor equations (\ref{Tmat}) we find explicit form of the
density and directional pressures as
\begin{align}\label{sol+}
 &\rho^+=\frac{\mu^2}{2\ell^2}\bigg\{\bigg(\frac{\bar{m}^2\tau^3}{5}\bigg)^2\exp\bigg(\frac{\bar{m}^2\tau^4}{10}-\tau^2\bigg)-\bar{m}^2\exp\bigg(\frac{\bar{m}^2\tau^4}{10}
 \bigg)\notag\\&~~~~-\bar{m}^2\exp\bigg(\frac{\bar{m}^2\tau^4}{10}
 +\tau^2\bigg)-\frac{\epsilon\mu^2}{4}\bigg[1+\bigg(\frac{\bar{m}^2\tau^4}{5}-1\bigg)\exp\bigg(\frac{\bar{m}^2\tau^4}{5}\bigg)\bigg]\bigg\}\notag\\&
 p_r^+=\frac{\mu^2}{2\ell^2}\bigg\{3\bigg(\frac{\bar{m}^2\tau^3}{5}\bigg)^2\exp\bigg(\frac{\bar{m}^2\tau^4}{10}-\tau^2\bigg)-\bar{m}^2\exp\bigg(\frac{\bar{m}^2\tau^4}{10}
 \bigg)\notag\\&~~~~-3\bar{m}^2\exp\bigg(\frac{\bar{m}^2\tau^4}{10}
 +\tau^2\bigg)-\frac{\epsilon\mu^2}{4}\bigg[1+\bigg(\frac{\bar{m}^2\tau^4}{5}-1\bigg)\exp\bigg(\frac{\bar{m}^2\tau^4}{5}\bigg)\bigg]\bigg\}\notag\\&
 p_t^+=\frac{\mu^2}{2\ell^2}\bigg\{5\bigg(\frac{\bar{m}^2\tau^3}{5}\bigg)^2\exp\bigg(\frac{\bar{m}^2\tau^4}{10}-\tau^2\bigg)-\bar{m}^2\exp\bigg(\frac{\bar{m}^2\tau^4}{10}
 \bigg)\notag\\&~~~~-3\bar{m}^2\exp\bigg(\frac{\bar{m}^2\tau^4}{10}
 +\tau^2\bigg)-\frac{\epsilon\mu^2}{4}\bigg[1+\bigg(\frac{\bar{m}^2\tau^4}{5}-1\bigg)\exp\bigg(\frac{\bar{m}^2\tau^4}{5}\bigg)\bigg]\bigg\} \end{align}
 for $+$ sign solutions and  \begin{align}\label{sol-}&\rho^-=\frac{\mu^2}{2\ell^2}\bigg[1+\bigg(\frac{\bar{m}^2}{2}\bigg)^2\exp\bigg(\bar{m}^2\tau-\frac{1}{\tau}
 \bigg)\notag\\&~~~~-\bar{m}^2\exp\bigg(\bar{m}^2\tau+\frac{1}{\tau}\bigg)+\frac{\epsilon\mu^2}{2}[2\bar{m}^2\tau-1]\exp(2\bar{m}^2\tau)\bigg]\notag\\& p_r^-=
 \frac{\mu^2}{2\ell^2}\bigg[1+3\bigg(\frac{\bar{m}^2}{2}\bigg)^2\exp\bigg(\bar{m}^2\tau-\frac{1}{\tau}
 \bigg)\notag\\&~~~~-3\bar{m}^2\exp\bigg(\bar{m}^2\tau+\frac{1}{\tau}\bigg)+\frac{\epsilon\mu^2}{2}[2\bar{m}^2\tau-1]\exp(2\bar{m}^2\tau)\bigg]\notag\\&
 p_t^-=
 \frac{\mu^2}{2\ell^2}\bigg[1+5\bigg(\frac{\bar{m}^2}{2}\bigg)^2\exp\bigg(\bar{m}^2\tau-\frac{1}{\tau}
 \bigg)\notag\\&~~~~-3\bar{m}^2\exp\bigg(\bar{m}^2\tau+\frac{1}{\tau}\bigg)+\frac{\epsilon\mu^2}{2}[2\bar{m}^2\tau-1]\exp(2\bar{m}^2\tau)\bigg] \end{align}
 for $-$ sign solutions respectively. In the following section we investigate physical properties of the obtained solutions above.
\section{Physical interpretations of the solutions}
  As we see at below that
 our model predicts a particular CW boson star with 'hard
surface' for which the matter density and radial pressure vanish
at a finite radius parameter $s$ defined by the boson mass (see
Eqs. (\ref{massp}) and (\ref{massm}) at below), but in many of
models, the boson stars have not a hard surface same as neutron
stars and density and directional pressures vanish asymptotically
at infinite distances (see for instance \cite{Betti}). Looking at
the figures 3-c and 4-a one infer that the directional pressures
vanish just for particular value of the matter density in case of
$+$ sign of the solutions and by comparing with the figures (3-d)
and (4-b) we obtain that this situation is not happened for $-$
sign solutions. Furthermore negativity values of the density in
figures (3-d) and (4-b) forces us to decide that the minus sign of
the solutions are not physical or it describes anti-matter
distribution of the star (say anti-star) with negative energies.
However we continue our statements just for solutions with $+$
sign solutions which at least give us some regular understandable
concepts. In this case prediction of a CW boson star with finite
scale can be considered an advantage and privilege of this current
work which come from CW self-interaction potential between the KG
bosons. In cases where the boson stars have not hard surface, one
usually use an estimated radius
\begin{align}\label{R} R=\frac{1}{Q}\int dx^3 \sqrt{g}rj^t=\frac{4\pi \omega}{Q}\int dr r^3 \phi^2 e^{\frac{H-U}{2}}\end{align}
in which $Q$ is Noether charge such that
\begin{align}\label{Jr}Q=-\int dx^3\sqrt{g}j^t=4\pi\omega\int r^2 dr \phi^2 e^{\frac{H-U}{2}},~~~j^t=\omega \phi^2 e^{-U}.\end{align}
In the above relation $j^t$ is the locally conserved current
associated to the globally conserved Noether charge $Q.$ In fact,
in the model with un-gauged U(1) symmetry, the Noether charge $Q$
is usually interpreted as the number of boson particles with mass
$`m`$ that make up the boson star. Usually, the scalar field
making up the boson star decays fast and so has not a hard
surface. In fact presence of the CW potential in this work causes
that the bosons do not decay fast and so surface of such a boson
star remains stable. Hence we do not apply to calculate estimated
radius (\ref{R}) of this kind of boson star but we determine exact
form of the radius parameter by solving the equations
$\rho^+(s)=0=p_r^{+}(s)$ for (\ref{sol+}) such that
\begin{equation}\label{massp}\bar{m}_+^2=\frac{5e^{s^2}}{s^3}\end{equation} with particular dimensionless CW potential coupling constant \begin{equation}
\label{epsilonp}\epsilon^+=\frac{4}{5\mu^2}\bigg[\frac{(1-s^2)e^{s^2}-s^2}{(se^{s^2}-1)\exp[\frac{s}{10}e^{s^2}]+\exp[-\frac{s}{10}e^{s^2}]}\bigg].\end{equation}
Substituting the above conditions into the density and pressures
given by the equations (\ref{sol+}) we find
\begin{align}\label{ptp}&\rho^+=\frac{\mu^2}{2\ell^2}\bigg\{\exp[s^2+se^{s^2/2}]-\frac{5\exp[s^2+se^{s^2}/2]}{s^3}-
\frac{5\exp[2s^2+se^{s^2/2}]}{s^3}\notag\\&~~~~-\frac{1}{5}\frac{[(1-s^2)e^{s^2}-s^2][1+(se^{s^2}-1)\exp[se^{s^2}]]}{(se^{s^2}-1)\exp[se^{s^2}/10]+\exp[-se^{s^2}/10]}
\bigg\}\notag\\&
p_r^+=\frac{\mu^2}{2\ell^2}\bigg\{3\exp[s^2+se^{s^2/2}]-\frac{5\exp[s^2+se^{s^2}/2]}{s^3}-
\frac{15\exp[2s^2+se^{s^2/2}]}{s^3}\notag\\&~~~~-\frac{1}{5}\frac{[(1-s^2)e^{s^2}-s^2][1+(se^{s^2}-1)\exp[se^{s^2}]]}{(se^{s^2}-1)\exp[se^{s^2}/10]+\exp[-se^{s^2}/10]}
\bigg\}\notag\\& p_{t}^+(s) =\frac{5\mu^2}{2\ell}
\exp\bigg(\frac{s}{2}e^{s^2}\bigg)
\bigg[\bigg(1-\frac{1}{s^3}\bigg)e^{s^2}-\frac{3e^{2s^2}}{s^3}\notag\\&
-\frac{\exp[-\frac{s}{2}e^{s^2}]}{25s^2}-\frac{e^{s^2+(s/2)\exp(s^2)}}{25s}+\frac{\exp[\frac{s}{2}e^{s^2}]}{25s^2}\bigg]
\end{align} which are defined versus the radius parameter $s.$\\
With same calculations, we can find position of radius of the CW
boson stars by solving $\rho^-(s)=0$ and $p_r^-(s)=0$ such that
\begin{align}\label{massm}\bar{m}_-=\exp{\bigg(\frac{1}{s}\bigg)}\end{align} with CW potential coupling constant \begin{align}\epsilon^-=\frac{2}{\mu^2}\bigg(
\frac{\exp[-8s e^{\frac{2}{s}}]}{1-8se^{\frac{2}{s}}}\bigg)
\end{align} and transverse pressure \begin{align}\label{pts}p_t^-(s)=\frac{4\mu^2}{\ell^2}\exp\bigg(\frac{3}{s}+4s e^{\frac{2}{s}}\bigg).\end{align}
By looking at the positive sign metric solution one infer that it
behaves as a Minkowski flat space time asymptotically at central
regions $\tau\to0$ while the metric solution with negative sign
treats same but at far from the central region $\tau\to\infty.$
Signature of both of the metric solutions are Euclidian (-,-,-,-)
which means that these metric solutions are for inside of the
star. We plotted  the mass functions (\ref{massp}) and
(\ref{massm}) versus the radius parameter $s$ in figures 2-a and
2-b respectively. Comparing them, one infer that there is a local
minimum point
 for $\bar{m}_+$ but not for $\bar{m}_-.$ Minimum value of the mass $\bar{m}_+$ is $\bar{m}^+_{min}=12.198$ at particular radius $s_p=\sqrt{1.5}=1.2248.$
 Physically, this minimum point describes that positive sign
  branch of the solutions tend to be remain as stable around this minimum point but the solutions with negative sign are not
  tend. Looking at the figure 2-b, one infer that the figure for $\bar{m}_-$ is absolutely decreasing function and so has not a local minimum point. It is confirmed
  by phase space trajectories given by figure 1-b where the system is in quasi-stable (saddle) nature. Because some of eigenvalues of the secular equation of the Jacobi matrix
  have positive and some other have negative sign.
  To understand more about physical behavior of the obtained solutions we plot figures of the re-scaled CW potential coupling constant
 $\bar{\epsilon}^+=\mu^2\epsilon^+/4$ and  $\bar{\epsilon}_-=2\epsilon/\mu^2$ versus the radius parameter $s$ in figures 2-c and 2-d respectively. \\
   Comparing the figures 2-c and 2-d we find that
effects of the CW self interaction of KG bosons are dominant for
$\bar{\epsilon}_+$ more than $\bar{\epsilon}_-$ at smaller scales.
Both of them have negative values.  Furthermore we plotted figures
of the dimensionless transverse pressures for both of the branches
of the solutions $\pm$ in figures 3-a and 3-b. Comparing them, one
can infer that the $+$ sign branch of the field solutions describe
dark stars with negative transverse surface pressure
but the $-$ sign branch of the field solutions show a visible boson star with positive transverse surface pressure.\\
One can check other ways to prove that our obtained solutions do
not describe black hole solutions but they show a star with
regular metric. From geometrical point of view, one usually solve
the horizon equation
$g^{\mu\nu}\partial_\mu\zeta(r)\partial_\nu\zeta(r)=0$ to
determine  radius of a black hole solutions where $\zeta(r)$ is a
spherical symmetric surface and it is surface of the black hole
horizon if a null vector field to be tangent to it. Using this
equation for the obtained solutions above gives us
$e^{-H_{\pm}}=0$ which by regarding the solutions (\ref{metsol})
one find solutions at $\tau=0$ and $\tau\to\infty$ which have not
physical meaning. This can be investigated by other way too as
follows: If we calculate values of the metric fields on the
surface of the CW boson star by substituting (\ref{massp}) and
(\ref{massm}) into the metric field solutions (\ref{metsol}), then
we find
\begin{align}e^{U_+(s)}=-\frac{\bar{\omega^2}}{5}s^3 e^{-2s^2},~~~e^{H_+(s)}=e^{s^2},~~~\phi^+(s)=\mu\exp\bigg(\frac{s}{4}e^{s^2}\bigg)\end{align}
and
\begin{align}e^{U_-(s)}=-\bar{\omega^2}e^{-\frac{3}{s}},~~~e^{H_+(s)}=e^{\frac{1}{s}},~~~\phi^-(s)=\mu\exp\bigg(\frac{s}{2}e^{\frac{2}{s}}\bigg).\end{align}
Obviously, one can see that none of them becomes zero or infinite
for a given value of radius $s$, which means that the found
stellar object is indeed a boson star with a non-singular metric.
They do diverge to infinity just for $s=0$ and $s\to\infty$ which
have not physical meaning. At last to study equation of state of
the system we plot $p_r^{\pm}$ versus the density functions
$\rho^{\pm}$ given by (\ref{ptp}) and (\ref{pts}) in figures 3-c
and 3-d. One can infer that the figure 3-c shows a physical system
in which for smaller densities the radial pressure is negative and
describes a collapsing object but for larger densities it is an
expanding unstable object because of positivity of the radial
pressure. There is just a particular density for which the $p_r^+$
takes a zero value describing a CW boson star with finite scale.
Matter content of a star can be described also by the barotropic
index defined by slop of the equation of state. In this way one
can look at the figures 3-c and 4-a, then who find a minimum
density as $\rho_{min}^+\approx20000\times(\frac{\mu^2}{2\ell^2})$
with corresponding pressures
$p^+_{r~min}\approx-35000\times(\frac{\mu^2}{2\ell^2})$ and
$p^+_{t~min}\approx-35000\times(\frac{5\mu^2}{2\ell^2})$ for which
slope of these figures is changed from negative to positive signs.
This predicts that for densities less than the minimum value
$\rho^+<\rho^+_{min}$ the CW boson star treats as a dark star
because of negativity of the slope (the barotropic index)
$\gamma_{t,r}=\frac{dp_{t,r}^+}{d\rho^+}<0,$ while for densities
larger than the critical one
 this star treats as
visible stellar object because of positive sign of the slope
$\gamma_{t,r}=\frac{dp_{t,r}^+}{d\rho^+}>0.$ This means that the
solutions with sign of $+$ can be both behavior as visible or
invisible (dark) stars depending on whether their density is
greater or less than this critical value. The right side figure
3-d has not physical content because of negativity values of the
matter density. Hence we exclude solution with negative sign as a
physical solutions and claim that this model gives out a physical
solution just with positive sign. We end this section by
investigating the anisotropy property of the CW boson star. This
is done by plotting the subtraction of the pressures $p_t-p_r$
called as anisotropy factor given by (\ref{ptp}) and (\ref{pts})
versus the corresponding density functions in figures 4-c and 4-d.
They show that the anisotropy is non-vanishing in both of $\pm$
branches of the solutions throughout the different scales of the
densities.
\section{Summary and outlook}
 In this work, we used a massive complex KG time dependent scalar field in the presence of
 self-interaction
 CW potential, to solve the Einstein metric equations in a spherically
  symmetric static form. The analysis reveals that the stability and formation of
  the boson star are significantly influenced by the self-interaction potential, which plays
   a pivotal role in the scalar field's behavior. Particularly this kind of potential makes a finite value of the radius of the
   star with hard surface which in usual ways a boson star has not this finite radius. We solved field equations by using dynamical systems
   approach in which stability of the system is investigated by determining sign of the eigenvalues of the Jacobi matrix of the field
   equations. Our solutions show a CW boson star which is
   anisotropic with finite scale. This kind of star is dark invisible or visible, depending on value of its density which whether is less or larger than
    the critical density.  The critical density is defined by the minimum value of the CW potential and a ultraviolet cut-off length scale which we consider
    to remove divergencies of radiative  corrections of Feynman diagrams.
     This statement is found by investigating the slope of the pressure-density figures in both of radial and transverse pressures.
   Although we solved small scale regime of the field equations
    because a boson star should be a compact stellar object but there is
    different behavior for the field equations at large scales of the boson matter distribution which we do not seek this term at the present work.
    As an extension of this work
    we like to investigate the latter  problem in our future work.
     \section{Acknowledgment}
     \textbf{Funding}\\
 The authors did not receive support from any
organization for the submitted work.\\
 \textbf{Ethics declarations}\\
 \textbf{Conflict of interest}\\
The authors have no competing interests to declare that are
relevant to the content of this article.

\begin{figure}
\centering \subfigure[{}]{\label{CW}
\includegraphics[width=0.45\textwidth]{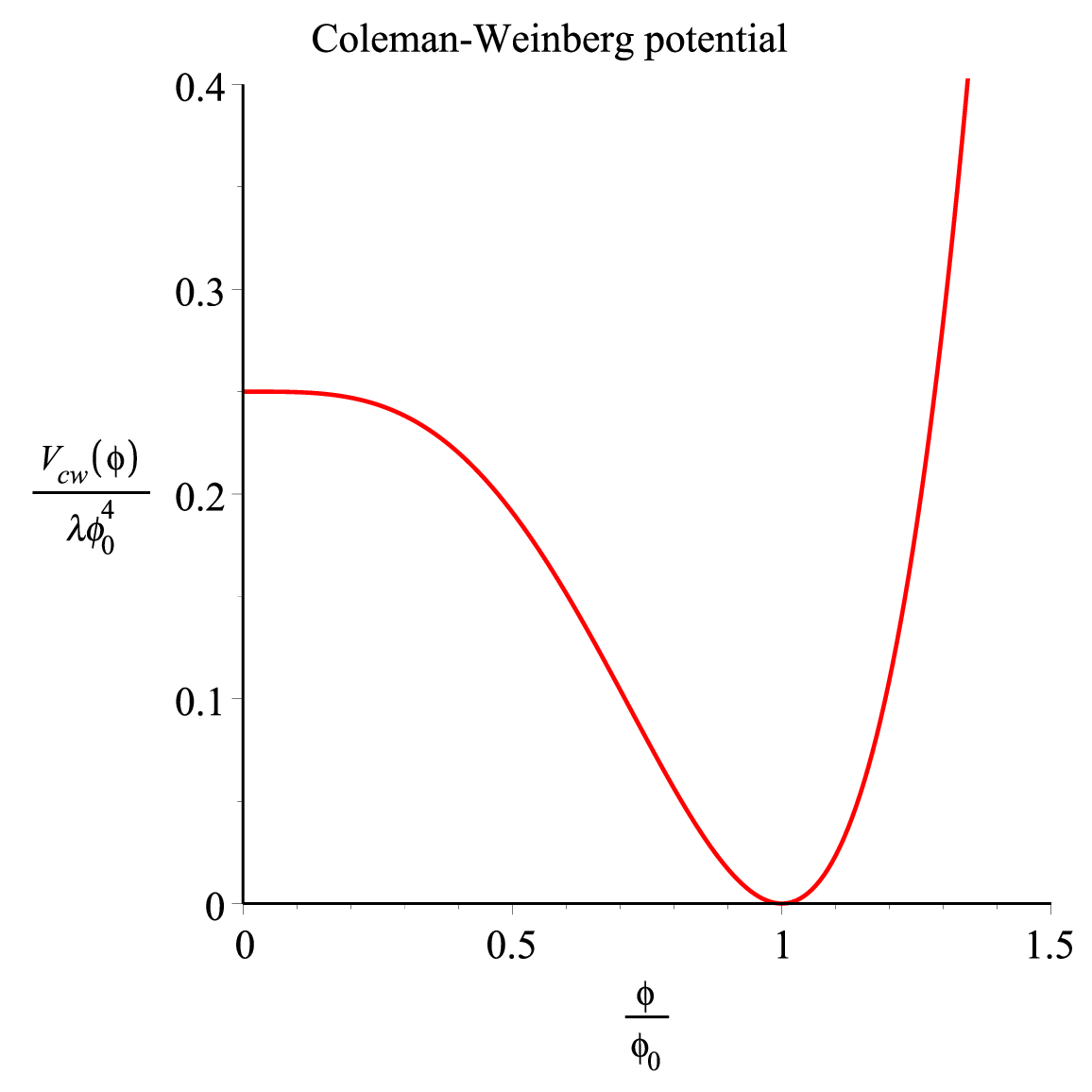}}
\hspace{1mm}\subfigure[{}]{\label{pr}
\includegraphics[width=0.45\textwidth]{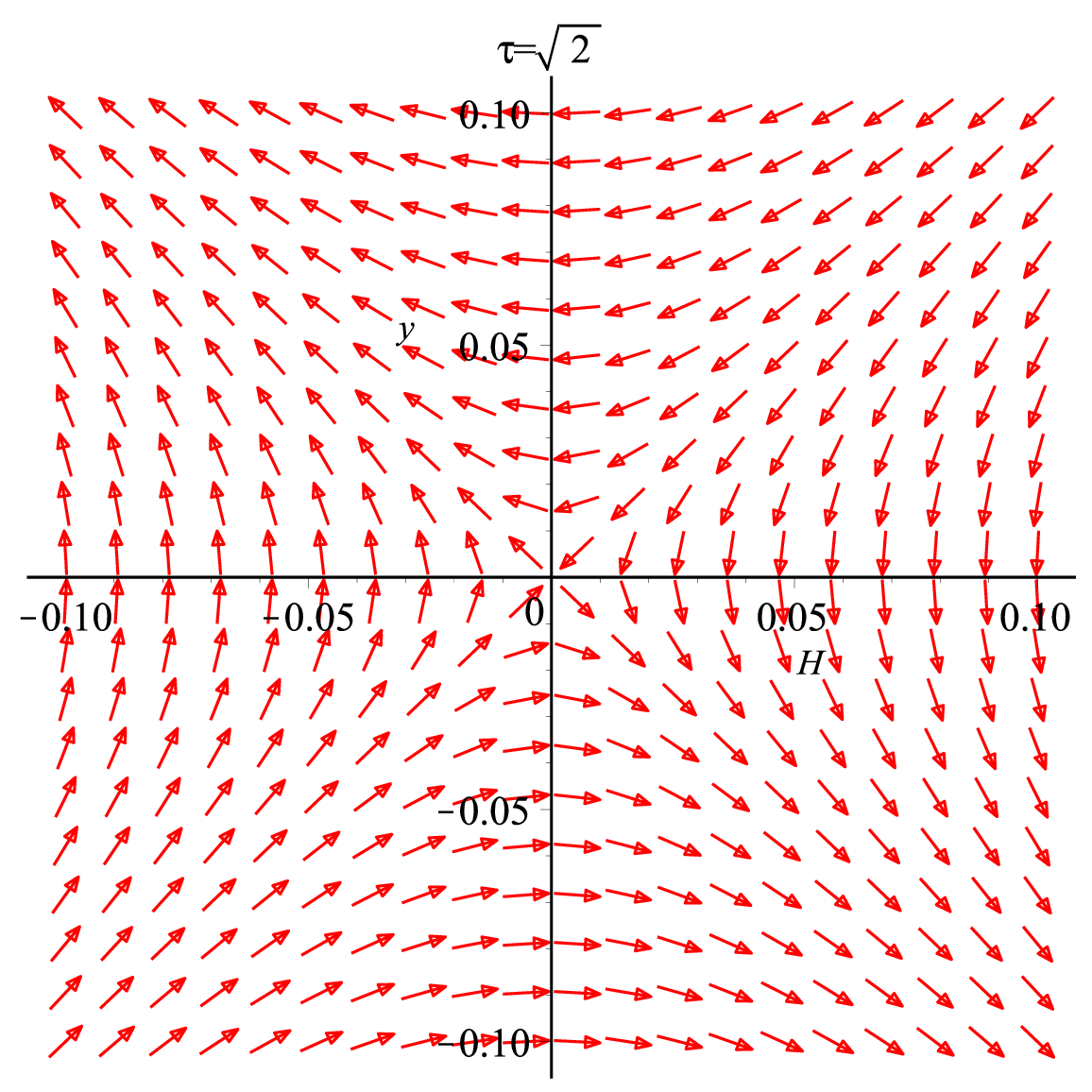}}
\hspace{1mm} \caption{\footnotesize{ (a) The Coleman-Weinberg
potential shows a minim value at $\phi_0=\mu$ which our solutions
are valid near this minimum point, (b) Phase space trajectories
show a quasi stable (saddle) nature for obtained solutions.} }
\end{figure}
\begin{figure}
\centering \subfigure[{}]{\label{x34001}
\includegraphics[width=0.45\textwidth]{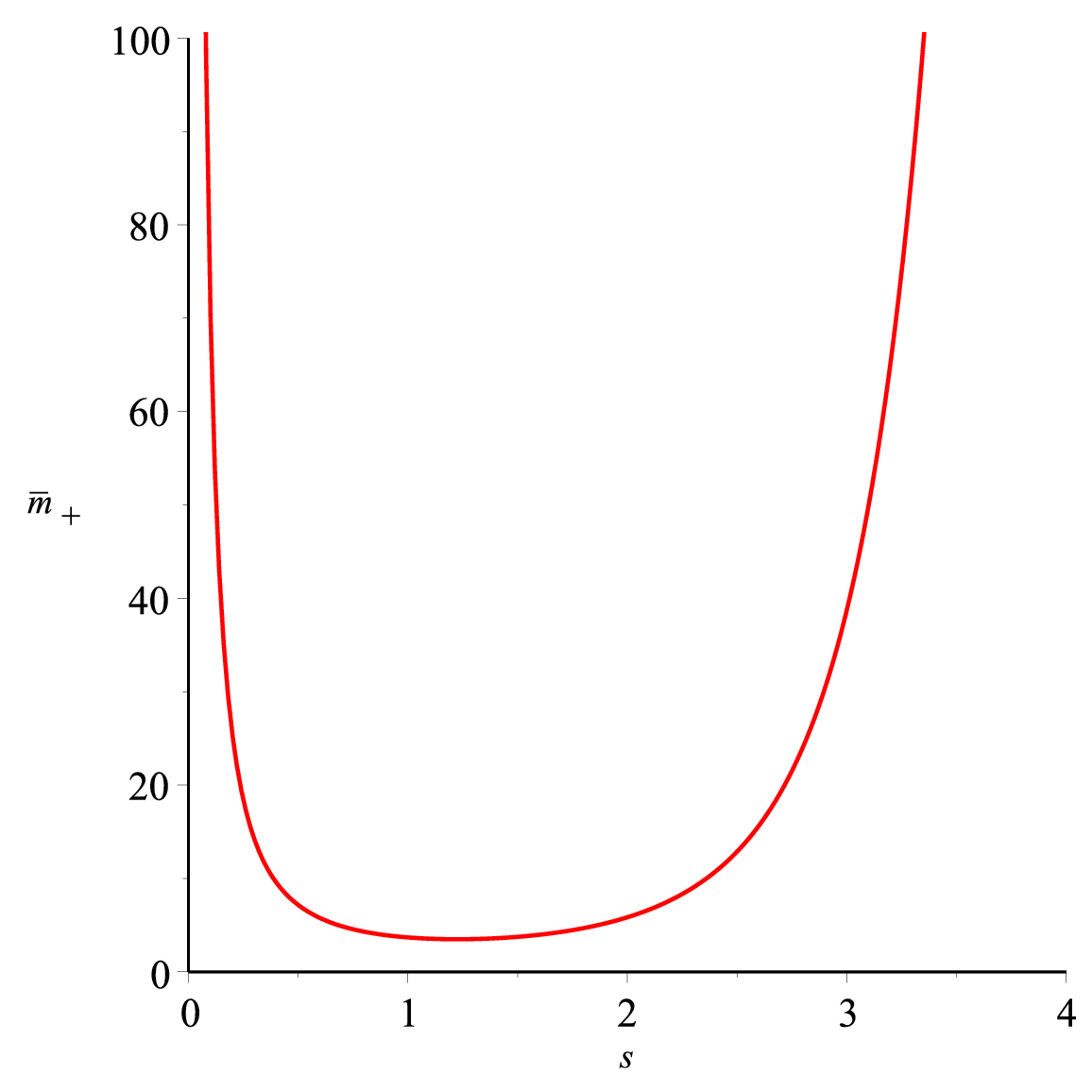}}
\hspace{2mm}\subfigure[{}]{\label{U2}
\includegraphics[width=0.45\textwidth]{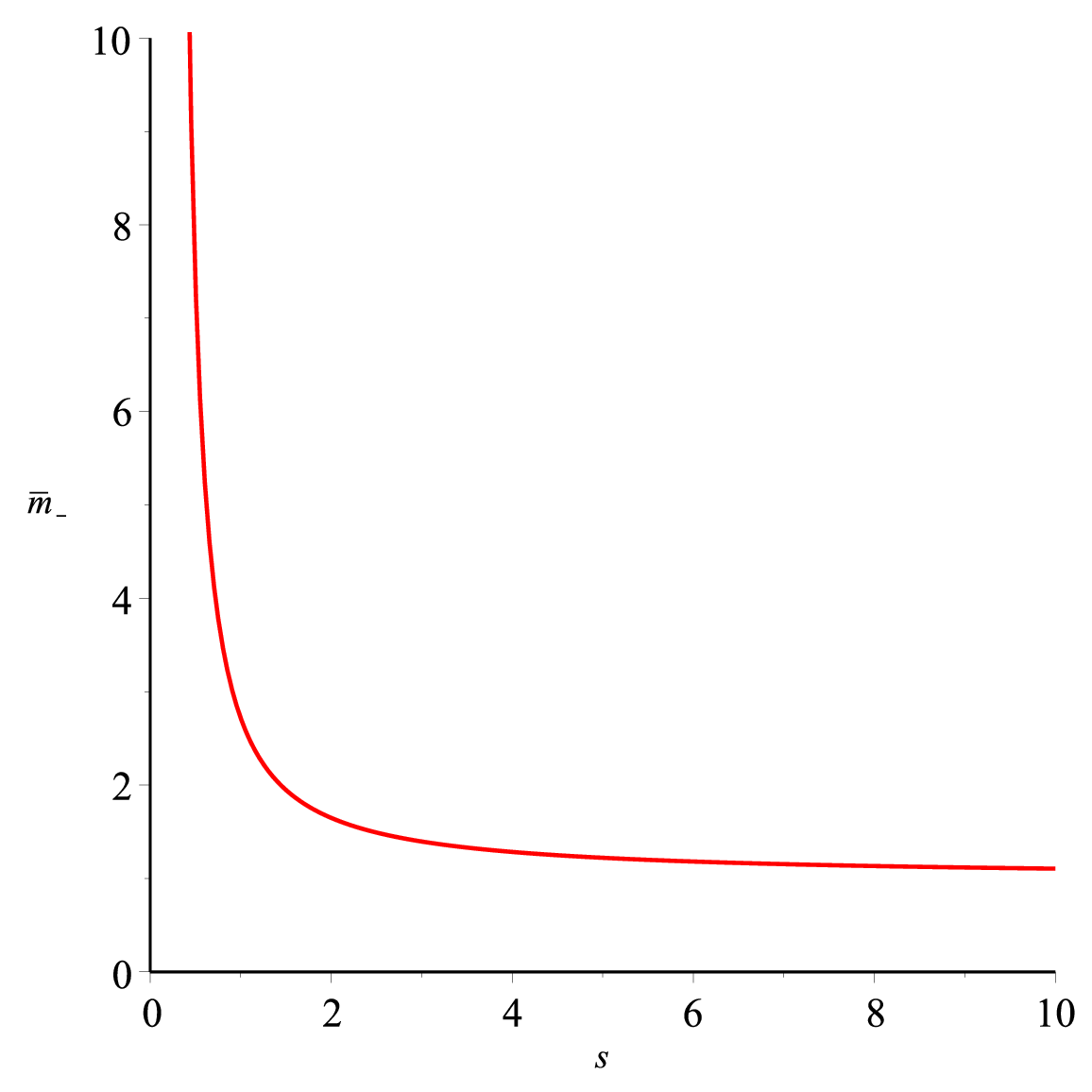}}
\hspace{2mm}\subfigure[{}]{\label{x34099}
\includegraphics[width=0.45\textwidth]{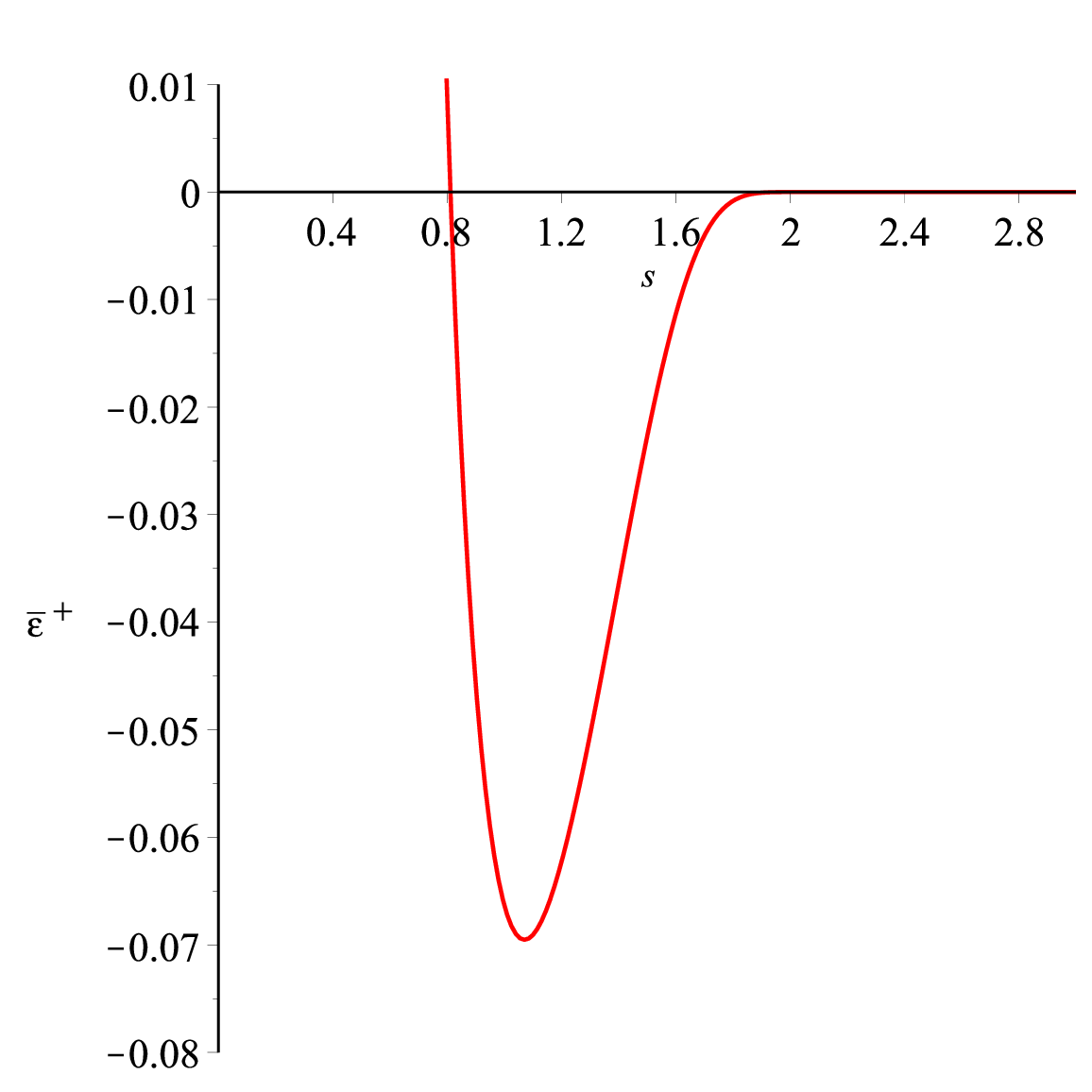}}
\hspace{2mm}\subfigure[{}]{\label{x34110}
\includegraphics[width=0.45\textwidth]{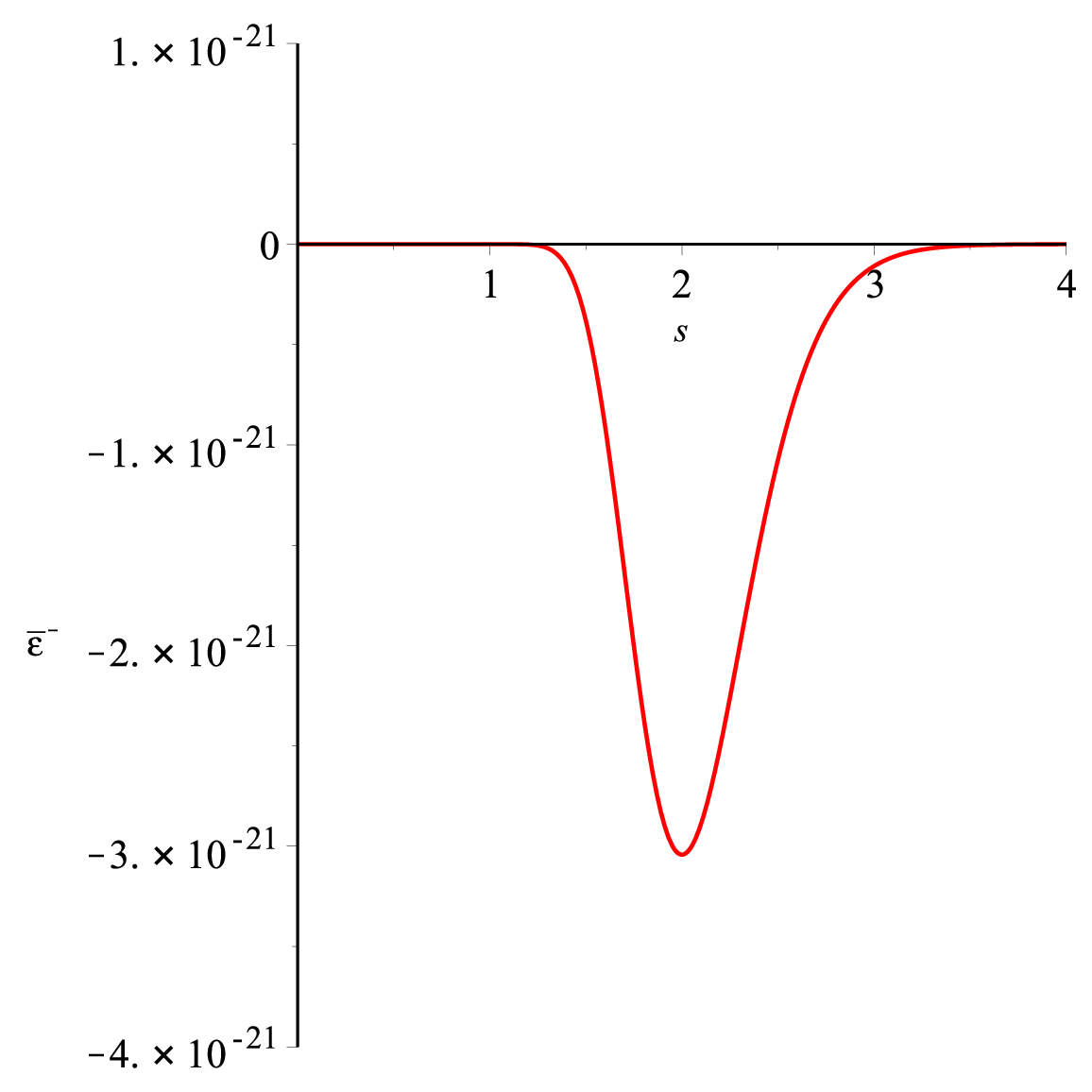}
} \hspace{2mm} \caption{\footnotesize{
 (a) mass parameter of the bosons
is plotted versus the radius parameter for $+$ sign solutions
where the mass parameter takes a local minimum value at
$s\sim1.3,$ (b) mass parameter of the bosons is plotted versus the
radius parameter for $-$ sign solutions. This choice of the
solution has not a minimum point and so it dose not predict a
stable state for finite scale CW boson star, (c) CW potential
coupling constant is plotted versus the radius parameter of the CW
boson star $s$ for solutions with $+$ sign. It shows that the CW
potential coupling constant is dominant just for radiuses $s<2,$
(d) CW potential coupling constant is plotted versus the radius
parameter of the CW boson star $s$ for solutions with $-$ sign. It
shows that this coupling constant is dominant for radius $s=2$
only. }}
\end{figure}
\begin{figure}
\centering \subfigure[{}]{\label{x34001}
\includegraphics[width=0.45\textwidth]{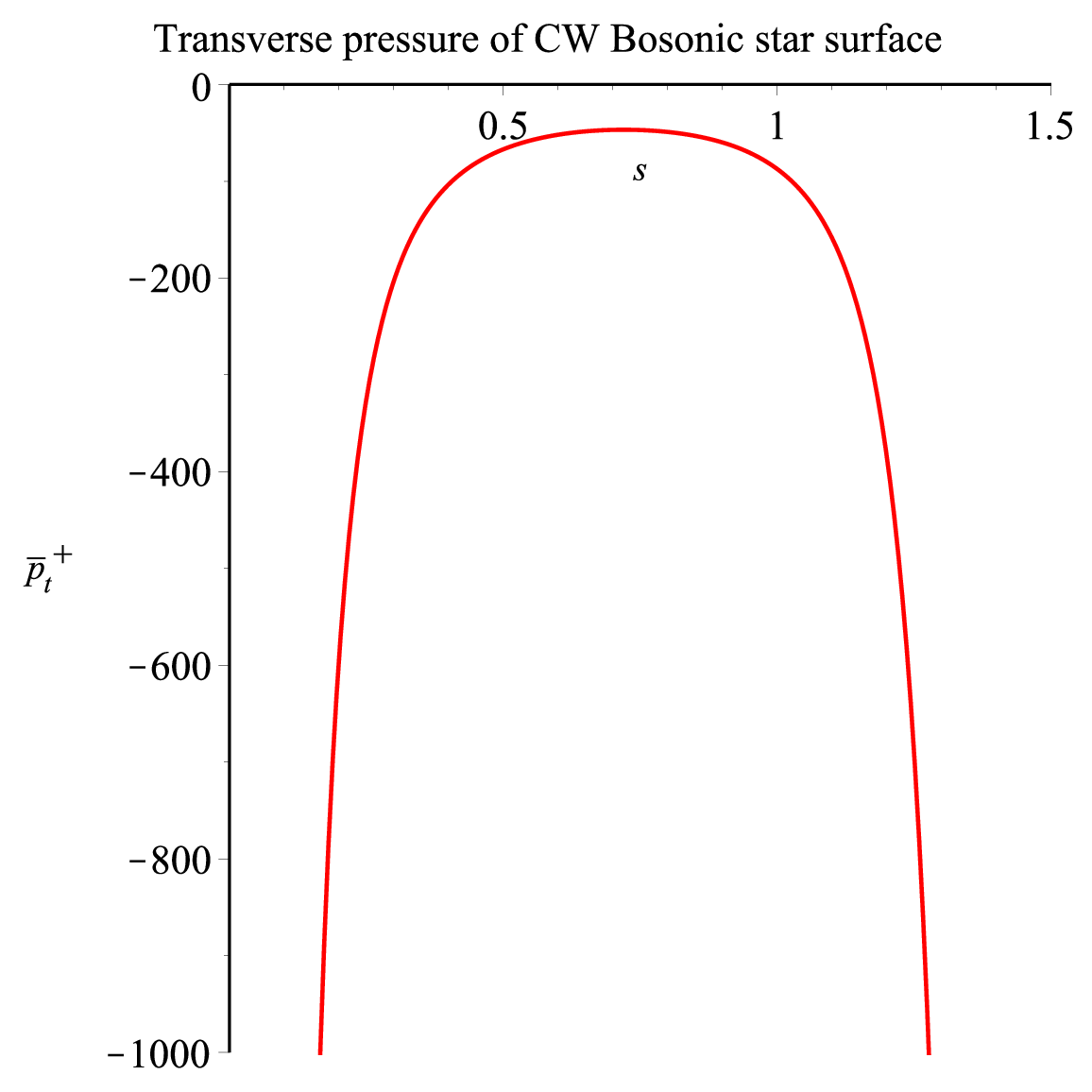}}
\hspace{2mm}\subfigure[{}]{\label{U2}
\includegraphics[width=0.45\textwidth]{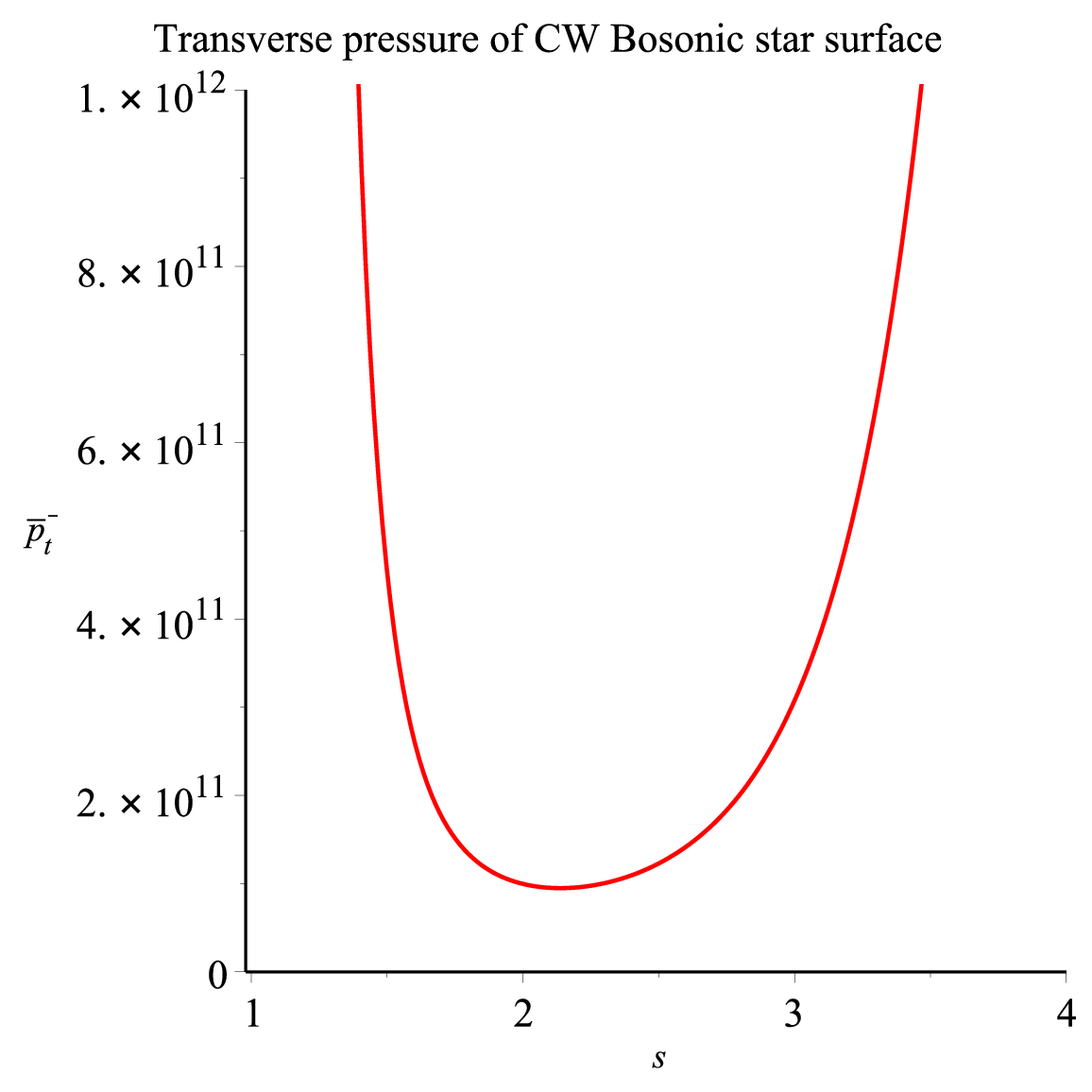}}
\hspace{2mm}\subfigure[{}]{\label{x34099}
\includegraphics[width=0.45\textwidth]{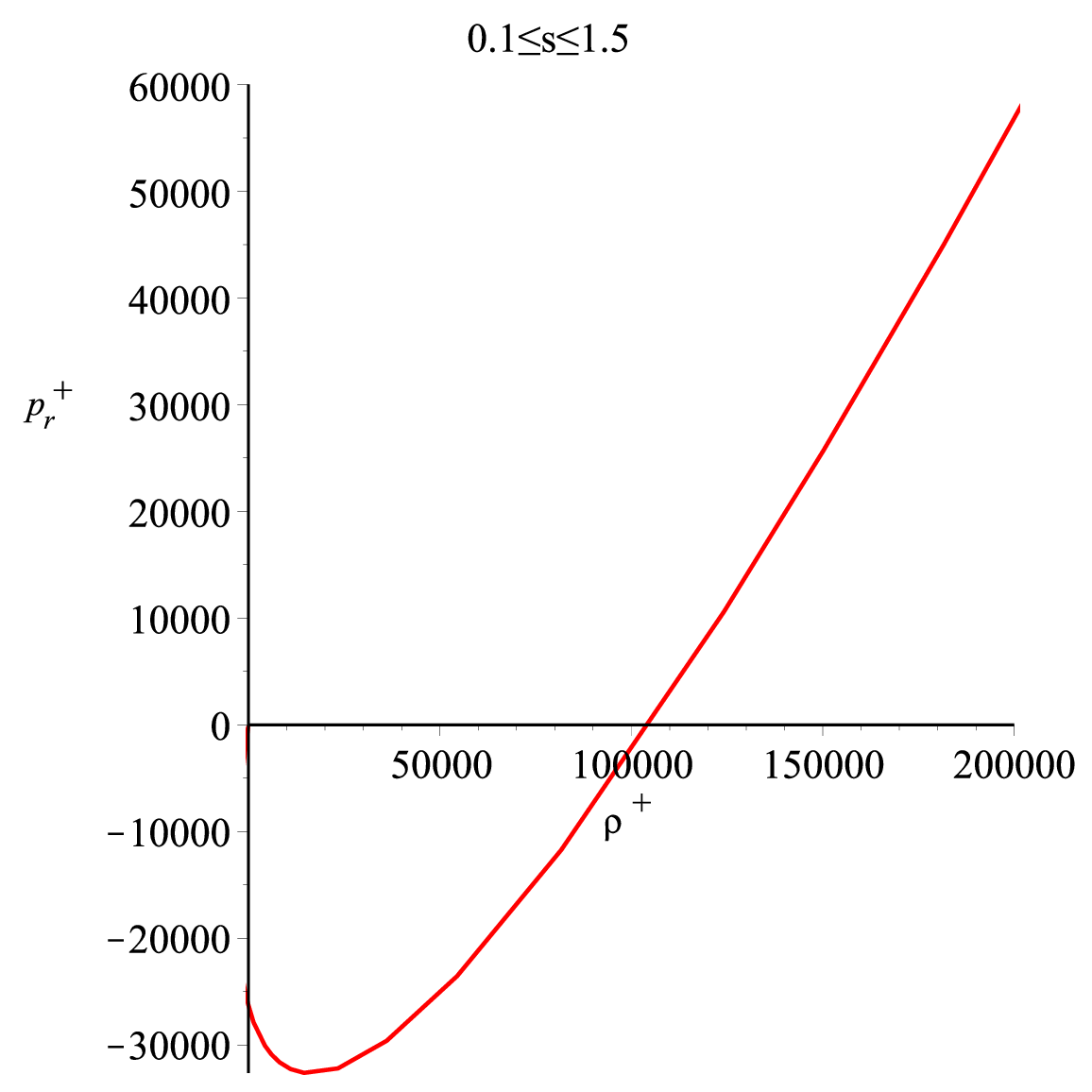}}
\hspace{2mm}\subfigure[{}]{\label{x34110}
\includegraphics[width=0.45\textwidth]{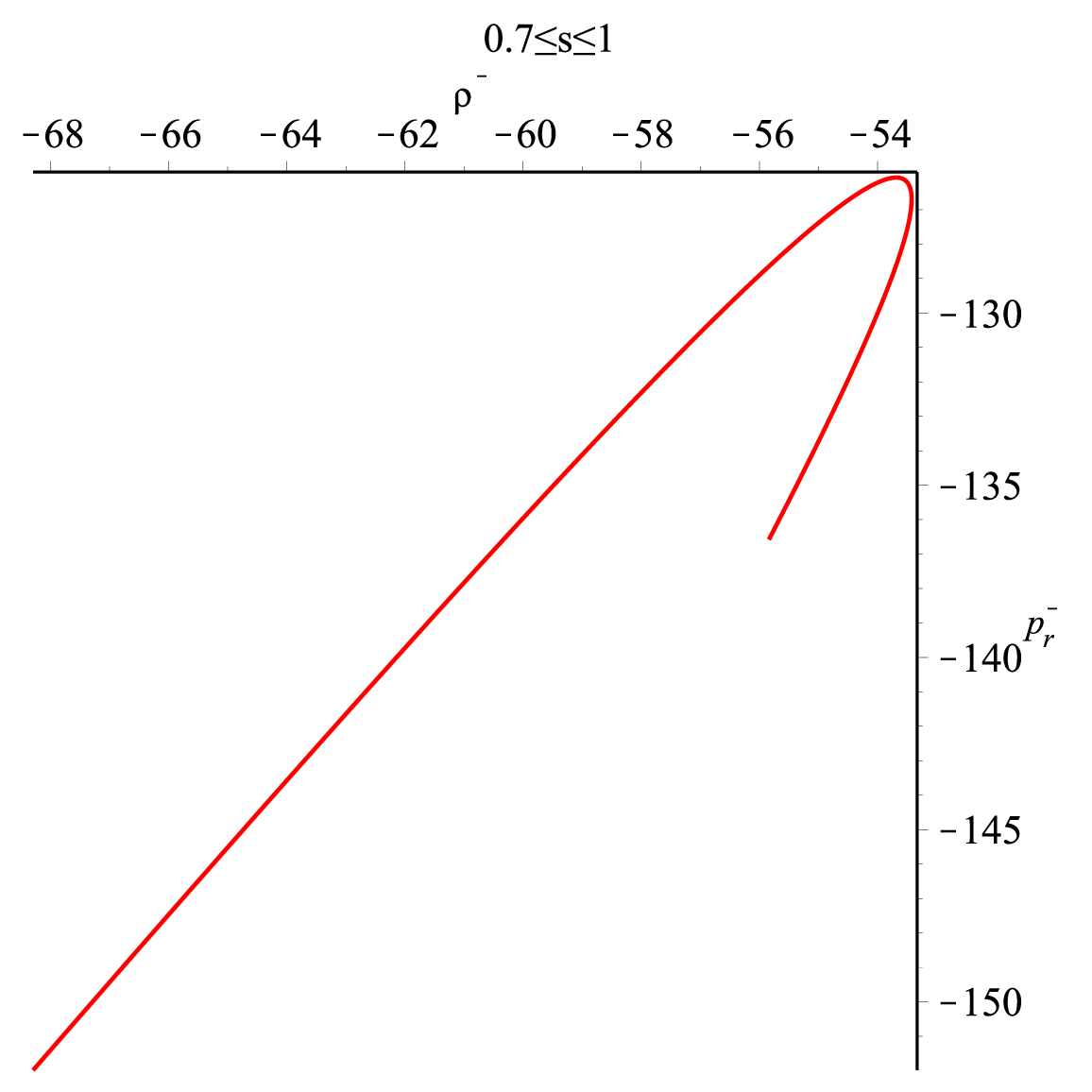}
} \hspace{2mm} \caption{\footnotesize{(a) shows variation of
transverse pressure versus the radius parameter $s$ for $+$ sign
solutions. Absolute negative values of this pressure can be able
us to claim that the matter content of this star behaves as dark
star, (b) shows variation of the transverse pressure versus $s$
for negative sign solutions. Absolute positive values for this
pressure shows that this kind of solution describes a visible
star, (c) shows variation of the radial pressure versus the
matter/energy density in case of solutions with $+$ sign. It has a
local minimum point. Slope of the figure has negative sign for
densities less that the minimum value which means this regime is
same as dark star but for densities larger than the minimum
density, the slope has positive sign showing visible phase of the
star (d) has not physical content because the density takes
absolutely negative values. }}
\end{figure}
\begin{figure}
\centering \subfigure[{}]{\label{x34001}
\includegraphics[width=0.45\textwidth]{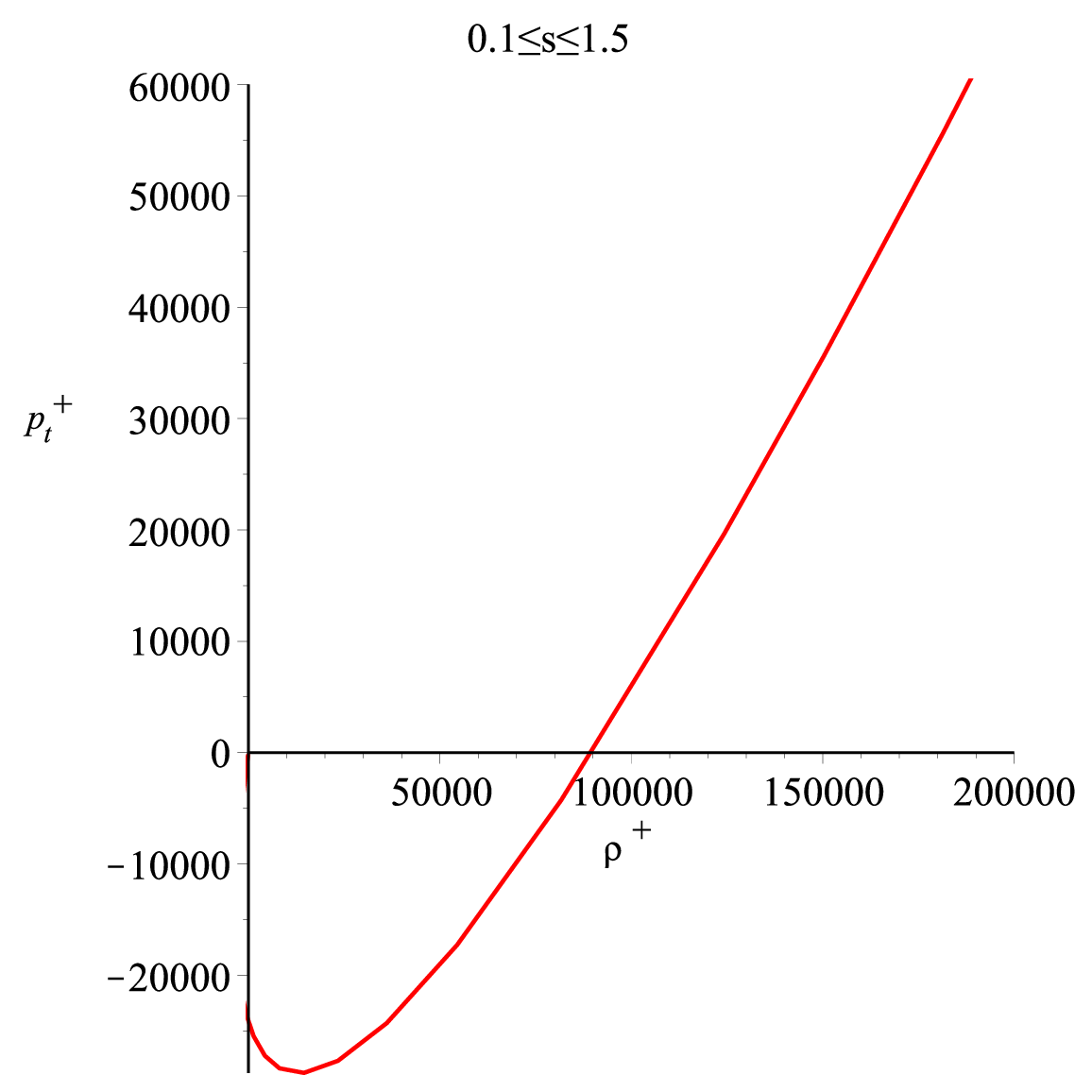}}
\hspace{2mm}\subfigure[{}]{\label{U2}
\includegraphics[width=0.45\textwidth]{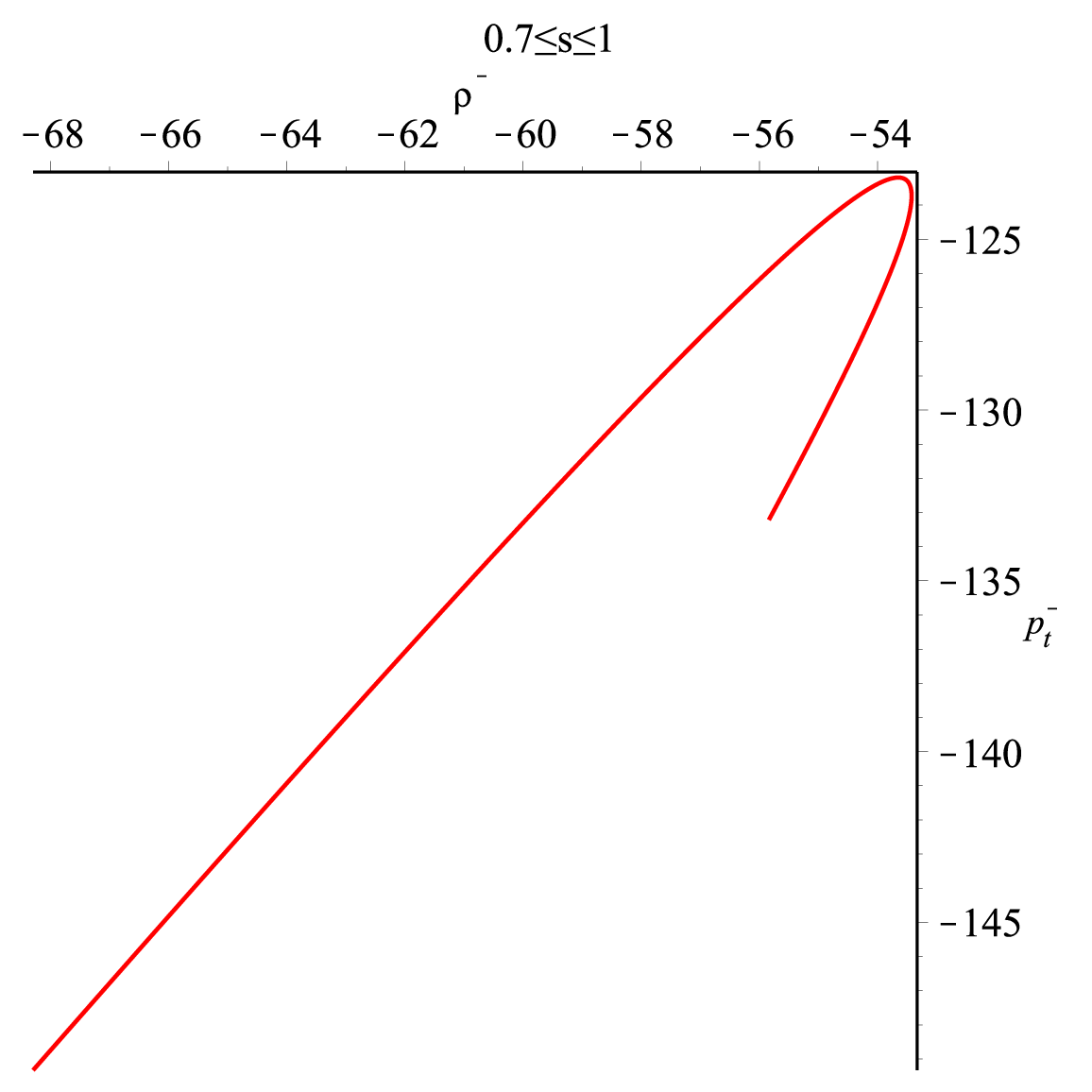}}
\hspace{2mm}\subfigure[{}]{\label{x34099}
\includegraphics[width=0.45\textwidth]{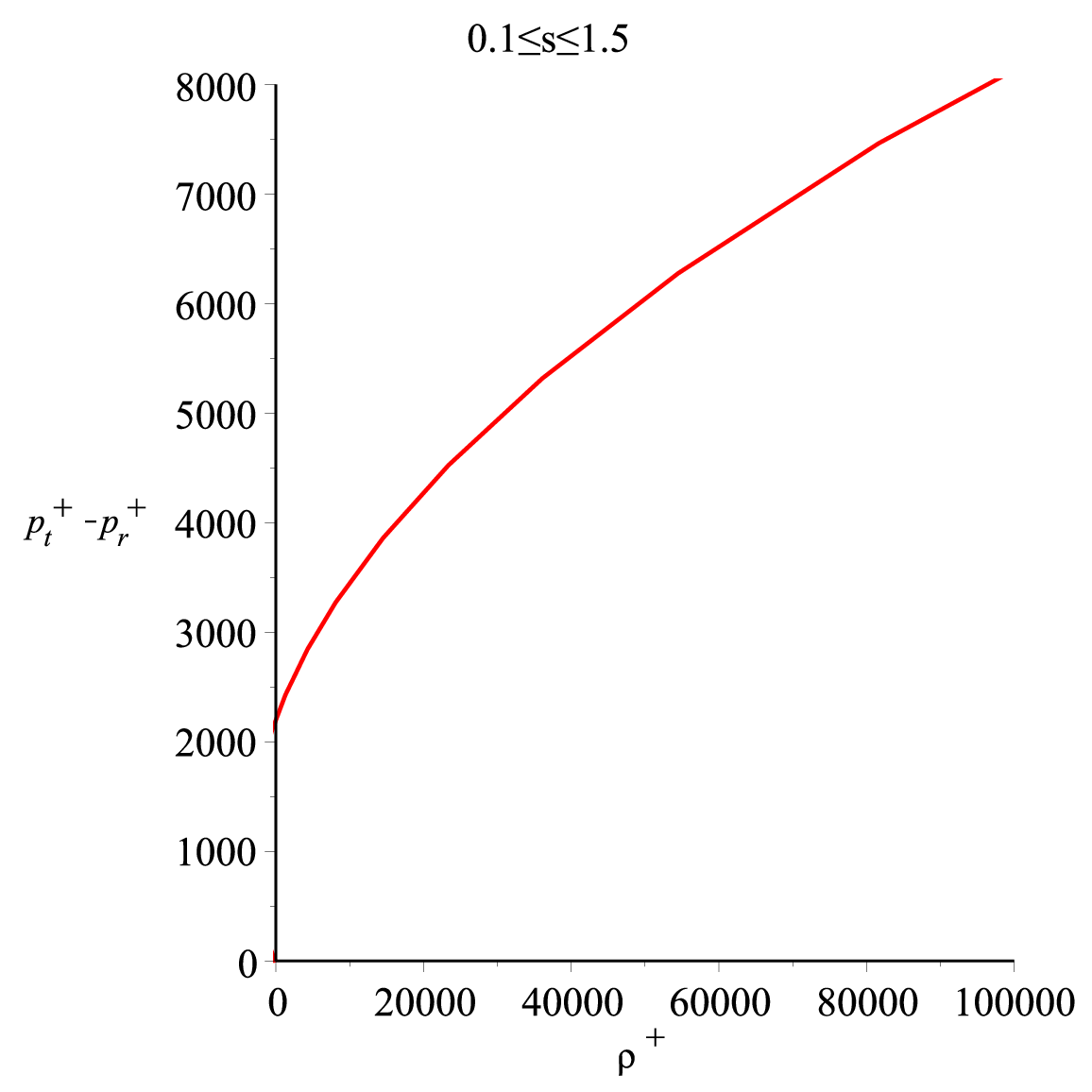}}
\hspace{2mm}\subfigure[{}]{\label{x34110}
\includegraphics[width=0.45\textwidth]{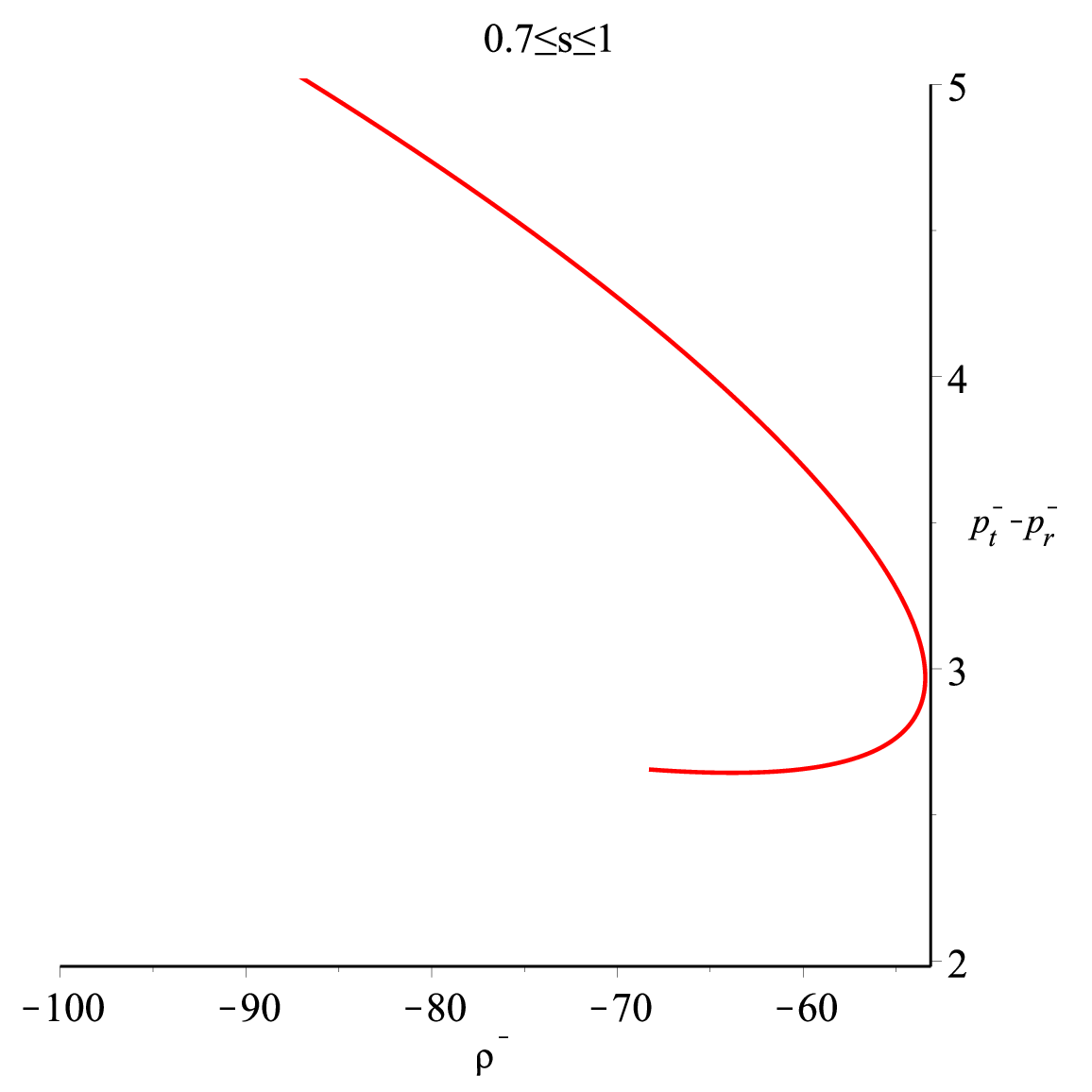}
} \hspace{2mm} \caption{\footnotesize{(a) shows variation of
transverse pressure versus the matter/energy density for $+$ sign
solutions. It has a local minimum point. Slope of the figure has
negative sign for densities less than the minimum value which
means this regime is same as dark star but for densities larger
than this minimum density the slope has positive sign showing
visible phase of the star, (b) has not physical content because
the density takes absolutely negative values for $-$ sign
solution, (c) shows anisotropy of the CW boson star in all scales.
(d) has not physics content because of negativity of the
density.}}
\end{figure}

\begin{thebibliography}{99}
\bibitem{sag} I. Sagret, M. Hempel, C. Greiner, and J. S. Bielich, Euro. J. Phys. 27, 577
(2006).
\bibitem{Jit} J. Kumar and P. Bharti, 2112.12518 [gr-qc]
\bibitem{con} J. J. O. Connor and E. F. Robertson, General relativity, Mathematical Physics index, School
of Mathematics and Statistics, University of St. Andrews, Scotland
(1996)
\bibitem{sco} S. H. Hawley, and M.W. Choptuik, Phys .Rev.D62,
104024 (2000); gr-qc/0007039
\bibitem{MM4} C. Vitor and
P. Paolo,  Living Rev Relativ, 22,4 (2019); 1904.05363v3 [gr-qc].
\bibitem{MMM4} P. A. Cano and L. Machet and
Ch. Myin,  2311.03433v2 [gr-qc]

\bibitem{stev} S. L. Liebling and C. Palenzuela, Living Rev. Relativity 26,
1(2023); 1202.5809 [gr-qc]
\bibitem{colp} M. Colpi, S.L. Shapiro and I. Wasserman, Phys. Rev. Lett. 57,
2485 (1986).

\bibitem{M1} Ph. Grandclement and S. Claire
and E. Gourgoulhon,  Phys. Rev. D90 024068, (2014); 1405.4837
[gr-qc]

\bibitem{Ghf} H. Ghaffarnejad, T. Ghorbani and F. Eidizadeh, Gen. Relativ. Gravit.55,135
(2023); 2301.00682 [gr-qc]
\bibitem{Ley} H. Ghaffarnejad and L. Naderi, Iran. J. Astro. Astrophys., Vol. 10,  365
(2023);2212.09485 [gr-qc]
\bibitem{vic} V. Jaramillo and  D. Nunez, Phys. Rev. D 106,
104023, (2022); 2209.07549 [gr-qc]
\bibitem{Font} D. S. Fontanella, and A. Cabo, 2101.04681 [gr-qc]
\bibitem{Hua} H. K. Guo, K. Sinha, C. Sun, J. Swaim and D. Vagie,
JCAP,0,028 (2021); 2010.15977 [astro-ph.CO]
\bibitem{wei} E. J. Weinberg, Ph.D. thesis,    hep-th/0507214
\bibitem{Hig} P.
Higgs, Phys. Letters 12, 132 (1964);
\bibitem{dan} D. Naegels, 2110.14504 [hep-th]
\bibitem{Mukh} V. Mukhanov, \textit{Physical foundations of
cosmology} Cambridge university press, (2005)
\bibitem{M2} F. Antonsen, K. Bormann   ,
hep-th/9702009v2
 \bibitem{M3} B. Nilay and
G. Amer and A. V. Nefer, JCAP, 1805, 046 (2018); 1802.04160v2
[astro-ph.CO]
\bibitem{M4} A. V. Nefer and Sh. Qaisar, Phys.Lett.B668:6, (2008);
hep-th/0806.2798v3
 \bibitem{M5} A. Afanasev et al, EPJA,60, 91 (2024)
\bibitem{M7} S. R. Coleman and E. J. Weinberg, Phys. Rev. D7,  1888 (1973).
 \bibitem{Liao} J. H. Liao, S. P. Miao and R. P. Woodard, Phys. Rev. D 99, 103522 (2019), 1806.02533v2 [gr-qc]
\bibitem{Betti}
Y. Brihaye,  L. Ducobu and B. Hartmann, Physics Letters B 811
135906 (2020)
\bibitem{Wein} S. Weinbrg, \textit{Gravitation and cosmology, Principar and applications of the general theory of
relativity}, John Wiley and Sons (1972).
\bibitem{cat} C. Catton, T. Faber and M. Visser, Class.
Quant. Grav.22:4189 (2005): gr-qc/0505137.
\bibitem{Hob} M. P. Hobson, G. P. Erstathiou and A. N. Lasenby \textit{"General
Relativity"
An Introduction for Physicists}, Cambridge University Press
(2006).

\bibitem{lav} P.Lawrence,\textit{Differential equations and dynamical systems} Springer, ISBN: 964-6213-52-9 (2001)
\bibitem{GHY} H. Ghaffarnejad, E. Yaraie, Gen. Relativ. Gravit.49, 49
(2017).



\end{thebibliography}
\end{document}